\documentclass[a4paper,fleqn]{mnras}

\usepackage{graphicx}	
\usepackage{amsmath}	
\usepackage{amssymb}	
\usepackage{pdflscape}	

\title{Neutrino Signals of Core-Collapse Supernovae in Underground Detectors}

\author[S. Seadrow et al.]{Shaquann Seadrow,$^{1}$\thanks{E-mail: shaquannseadrow@gmail.com}
Adam Burrows,$^{1}$
David Vartanyan,$^{1}$
David Radice$^{1,2}$
\newauthor
and M. Aaron Skinner$^{3}$
\\
$^{1}$Department of Astrophysical Sciences, Princeton University, Princeton, NJ 08544\\
$^{2}$Schmidt Fellow, Institute for Advanced Study, 1 Einstein Drive, Princeton, NJ 08540\\
$^{3}$Livermore National Laboratory, 7000 East Ave., Livermore, CA 94550-9234
}

\date{Accepted XXX. Received YYY; in original form ZZZ}

\pubyear{2017}

\begin{document}
\label{firstpage}
\pagerange{\pageref{firstpage}--\pageref{lastpage}}
\maketitle

\begin{abstract}
For a suite of fourteen core-collapse models during the dynamical first 
second after bounce, we calculate the detailed neutrino ``light" curves expected in the underground 
neutrino observatories Super-Kamiokande, DUNE, JUNO, and IceCube.  These results are
given as a function of neutrino-oscillation modality (normal or inverted hierarchy) and 
progenitor mass (specifically, post-bounce accretion history), and illuminate the 
differences between the light curves for 1D (spherical) models that don't explode with 
the corresponding 2D (axisymmetric) models that do. We are able to identify clear 
signatures of explosion (or non-explosion), the post-bounce accretion phase, and 
the accretion of the silicon/oxygen interface.  In addition, we are able to estimate 
the supernova detection ranges for various physical diagnostics and the distances out 
to which various temporal features embedded in the light curves might be discerned. 
We find that the progenitor mass density profile and supernova dynamics during the 
dynamical explosion stage should be identifiable for a supernova throughout most of 
the galaxy in all the facilities studied and that detection by any one of them, 
but in particular more than one in concert, will speak volumes about the internal dynamics 
of supernovae.  
\end{abstract}


\section{Introduction}
\label{intro}

Neutrinos are produced and radiated at very high rates during 
and immediately after the collapse of the core of a massive star ($M \ge 8$ M$_{\odot}$)
at its terminal stages (Burrows, Klein, \& Gandhi 1992; Fischer et al. 2009; 
M\"uller \& Janka 2014; Nakamura et al. 2016). For seconds 
during its dynamical and supernova phases, the core transitions 
into a proto-neutron star (PNS; Burrows \& Lattimer 1986), its neutrino emissions 
directly reflecting in real time the internal dynamics of this violent 
phenomenon. Theory identifies neutrinos as the drivers of explosion (Bethe \& Wilson 
1985) and of the evolution of the nascent PNS. As such, neutrinos are
the primary diagnostics of all stages of core behavior after stellar death
and before the emergence of the residual neutron star or black hole.
Stamped on the associated neutrino light curves, spectra, and 
mix of emitted neutrino species as a function of time are the 
distinctive signatures of the collapse, bounce, shock breakout, accretion, 
explosion, and PNS cooling phases, otherwise obscured from direct 
scrutiny by the profound opacity to photons of the overlying stellar
matter (Burrows 1990; Burrows, Hayes, \& Fryxell 1995; Fischer et al. 2009; 
M\"uller \& Janka 2014; Mirizzi et al. 2016; Nakamura et al. 2016; Horiuchi 
\& Kneller 2017). By the high signal rate, followed by a abrupt cessation,
one can witness in the neutrino emission history the formation,
if it occurs, of a black hole (Burrows 1984,1988; Sumiyoshi et al.
2006; Fischer et al. 2009).  The shock breakout burst in electron neutrinos
is a firm prediction of theory that must be tested (Thompson, Burrows, \& Pinto 2003;
Wallace, Burrows, \& Dolence 2016). The rapid neutrino light-curve variations due 
to turbulence in the core can constrain the convective physics predicted in
most multi-dimensional models (Lund et al. 2010,2012; Tamborra et al. 2013)
and considered essential to the generic mechanism of explosion (Herant et al. 1992; 
Burrows, Hayes, \& Fryxell 1995; Janka \& M\"uller 1996). 

The supernova neutrino burst has been detected only once, from the supernova SN1987A
at $\sim$50 kiloparsecs (kpc) in the large Magelanic Cloud (Hirata et al. 1987; 
Bionta et al 1987). However, a total of only 11 and 8 events over $\sim$12.5 
and $\sim$5.5 seconds, respectively, were culled.  While the crude basics of
PNS formation and cooling were verified in broad outline (Burrows \& Lattimer 1986,1987),
the signal rates in these relatively small detectors at this extragalactic distance
were paltry and most of the predicted phenomena were grossly undersampled. 
However, with the advent of numerous large-volume/mass underground detectors for the measurement
of solar and atmospheric neutrinos (Super-Kamiokande, Abe et al. 2016; IceCube, Abbasi et al. 2011, K\"opke et al. 2011)
and neutrino oscillation parameters (DUNE, Ankowski et al. 2016; JUNO, Lu et al. 2015),
a network of highly-capable facilities is being assembled that could record many 
thousands of neutrino events from a galactic core-collapse supernova (CCSN) explosion 
(Scholberg 2012; Beck et al. 2013).  Super-K ($\sim$32.5 kilotonnes detecting volume) 
may be followed by Hyper-K (Abe et al. 2011,2018; $\sim$0.25$-$0.4 Megatonnes) may be 
able to reach out to the Andromeda galaxy (M31) at $\sim$700 kiloparsecs (kpc).  
With such detectors, the detailed evolution of 
1) the neutrino spectra of the various neutrino types (Gallo Rosso et al. 2018), 2) the breakout burst, 
3) the variation in the mass accretion rate post-bounce, but 
prior to explosion, 4) the accretion of shell interfaces (such as that at 
the boundary of the silicon and oxygen shells), 5) the explosion itself, and 6)
the temporal fluctuations due to turbulent convection behind the shock 
could all be captured, identified, and studied. 

Hence, detecting such a high-signal galactic burst and its detailed scrutiny are
essential to validate and fundamentally constrain the theory of CCSN that has 
emerged over the last fifty years. The theory of the CCSN explosion and the theory of massive
star evolution to Chandrasekhar collapse (Woosley \& Weaver 1995; Woosley, Heger, \& Weaver 2002;
Hirschi, Meynet, \& Maeder 2004; Woosley \& Heger 2007; Maeder \& Meynet 2012;
Sukhbold et al. 2016) are intimately intertwined, so the latter is tested 
as well if a high-event rate galactic supernova is witnessed.

In this paper, we do not explore all the various neutrino signatures of the 
multi-second evolution of the supernova core, nor the retrieval in detail 
of the multiple parameters and inputs to CCSN theory.  This would be too 
large a task, though aspects of this will be relegated to future installments in 
our series on core-collapse signatures and diagnostics\footnote{Currently, in 
this series are Wallace et al. (2016) on the neutrino breakout burst and Morozova et al. 
(2018) on the gravitational wave emissions and characteristics.}. 
Rather, using recently generated two-dimensional (2D) and 
one-dimensional (1D) multi-group radiation/hydrodynamic models 
employing the code F{\sc{ornax}} (Burrows et al. 2018; Radice et al. 2017, Skinner 
et al. 2016; Wallace et al. 2016; Vartanyan et al. 2018; Skinner et al. 2018), 
we focus on the general overarching features of the neutrino light curve 
in Super-K, DUNE, JUNO, and IceCube during approximately the first second 
after bounce as a function of progenitor mass and between exploding and 
non-exploding models. The first second is the most dynamical phase. We also 
explore the potential identification of features in the accreted density structure, 
such as the silicon/oxygen interface, and estimate the effects of neutrino 
oscillations on the detected signals (Mirizzi et al. 2016; Scholberg 2018).  
We use the 9, 10, 11, 16, 17, 19, and 21 M$_{\odot}$ progenitor models of 
Sukhbold et al. (2016), as evolved in Radice et al. (2017), Burrows et al. 
(2018), and Vartanyan et al. (2018), for our representative progenitor model 
suite.  The 2D variants of these models explode (Vartanyan et al. 2018; Burrows et al. 2018; 
Radice et al. 2017), while the spherically-symmetric 1D models do not.  We use this 
difference to gauge the difference between exploding and non-exploding models
and find that after explosion the neutrino event rate is diagnostically very different
between them.  Also, we can identify in many of the non-exploding models a clear signature 
of the accretion of the silicon/oxygen interface.  Moreover, we see other characteristic 
timescales that can readily be discerned, as well as clear differences in all detectors 
between event-rate models in the various detectors for non-oscillation, normal-hierarchy,
and inverted-hierarchy realizations.

We emphasize that the differences between the 1D
and 2D models and between non-exploding and exploding models are not strictly related
to the specifics of our hydrodynamic models, but to the signature in underground 
detectors of the cessation of accretion power due to explosion.
This is a generic prediction of supernova theory and is not specific to 
our models.  In theory, the neutrino emissions from a supernova (or the 
collapsed core of a massive star) are powered by two sources - diffusion from
the core and accretion onto the core.  Given an initial density profile, reflected
in the initial progenitor model, the neutrino emissions after bounce and before explosion
are determined by this initial structure (given the physics of transport, equation of state, etc.).
Hence, there is a direct mapping between the progenitor structure (represented in the
literature and in this paper by the progenitor mass) and the pre-explosion neutrino
emissions. The onset of explosion inaugurates the cessation (and reversal) of accretion,
withdrawing this component from powering the emergent neutrino luminosity.  1D, 2D, and 3D
models of this emission before explosion are very closely the same, since the inner
core from which the neutrinos emerge is always pseudo-spherical.  This is seen before
explosion in the luminosity and signal plots provided in \S\ref{results}.  The multi-D effects are predominantly
turbulence behind the shock, shown in the literature to be crucial to explosion,
but still roughly averaging to spherical accretion (with some rapid temporal fluctuations),
as far as the emergent neutrino emissions are concerned. Turbulence introduces,
among other things, a turbulent pressure that aids and enhances ``explodability"
and brings the core to the critical condition for explosion.  

We reiterate that even if core collapse does not result in a supernova
explosion, the neutrino signal from such a core will be robust and observable. 
For our galaxy, non-exploding models have signal rates and total event counts that
are comparable to (and at later post-bounce times generally even larger than) those
for exploding models. The neutrino facilities will easily see this (if they are online),
even if there is no optical counterpart. It is the differences between exploding models
and non-exploding models (represented in this paper by 2D versus 1D models), due to
the cessation of accretion in the former, that we contend can be distinguished when
the next core collapse occurs in our galactic neighborhood.

Our purpose with this paper is not to rehash this theory, but to demonstrate
differences, in principle detectable in underground neutrino facilities, between exploding and
non-exploding models that are generic predictions of general core-collapse supernova
theory.  The prediction that there is a transition at explosion from accretion plus
core diffusion in the sourcing of the neutrino emissions to solely diffusion, and the
concomitant decrease at explosion and afterward of the neutrino signal(s), is an
experimentally testable hypothesis of current theory. In this paper, we provide
not just one model for signals, but a full suite, to quantify the expected approximate 
trends from low to higher mass supernova progenitors.  Most papers exploring signals provide
only one or two models, so doing the multitude of models we have presented here
is new to this paper.

In section \ref{signal_method}, we summarize our method of calculating a neutrino light curve
in the typical detector. We include in \S\ref{osc} a short discussion on incorporating
adiabatic neutrino oscillations.  We follow this in section \ref{detectors} with descriptions of the various 
underground neutrino detectors highlighted in this paper.  Then, in section \ref{snmodels},
we summarized the salient features of F{\sc{ornax}} and of the 2D and 1D multi-group 
radiation/hydro models generated using it, and follow in section \ref{results} with our results for the 
various neutrino light curves in Super-K, DUNE, JUNO, and IceCube.  Sections \ref{ranges} 
and \ref{accuracy} contains a general discussion of the implications of our results and 
section \ref{conclusions} summarizes our conclusions.  We postpone to another paper 
the investigation of the extraction of the neutrino energy spectrum and of the 
species mix of the neutrinos emitted.

\section{Signal Calculation Method}
\label{signal_method}

In this paper, we explore the characteristic neutrino signatures in various representative 
underground neutrino detectors during the first crucial and dynamical second after core bounce  
of the supernova phenomenon.  We investigate their dependence upon progenitor mass, distance, oscillation modality, 
and detector capabilities. The core-collapse models (\S\ref{snmodels}) were generated by our F{\sc{ornax}} code (Skinner et al. 2016; 
Radice et al. 2017; Burrows et al. 2018; Vartanyan et al. 2018) and we employ the SNoWGLoBES software package 
(Hubber et al. 2010; Beck et al. 2013; Scholberg 2018) for computing the detected event rates in the chosen constellation 
of underground detectors.  We have produced a pipeline that allows us easily to change the neutrino source, distance, and detector.

The neutrino energy spectra are calculated at 10,000 km from the center of each progenitor 
for models using either the LS220 EOS (equation of state) or the SFHO EOS. Since our focus 
is on the general signal characteristics of exploding vis \`a vis non-exploding
models (\S\ref{snmodels}), where all the 2D (axisymmetric) models we show explode 
and all the corresponding 1D (spherical) models do not, the specific EOS employed for each 
model is of secondary concern.  We lump together the $\nu_{\mu}$, $\bar{\nu}_{\mu}$, 
$\nu_{\tau}$, and $\bar{\nu}_{\tau}$ neutrinos into ``$\nu_x$" a neutrino bin, and 
distinguish the electron-type ($\nu_e$) and anti-electron-type ($\bar{\nu}_{e}$) species.
All three species have distinct spectra and temporal evolutions.  
SNOwGLoBES accounts for the multitude of interactions and channels in each detector caused by 
each species.   In an intermediate step required for signal determination, we convert 
the energy spectra,  $\frac{ dL_{\nu}\left(E_{\nu},t \right)}{dE_{\nu}},$ 
into number-luminosity spectra:
\begin{equation}
 \frac{d^2 N_{\nu}}{dE_{\nu} dt} =  \frac{ dL_{\nu}\left(E_{\nu},t \right)}{dE_{\nu}}  \frac{1}{E_{\nu}}\, .
\end{equation}
 
For a given detector, the event rate for any given channel is
\begin{equation}
\frac{dN_{det}}{dt} = N_{tar}  \frac{1}{4 \pi D^2} \int \frac{d^2 N_{\nu}}{dE_{\nu} dt} 
\sigma \left(E_{\nu} \right) \epsilon(E) dE_{\nu}\, ,
\end{equation}
where $N$ is the number of detected events, $N_{\nu}$ is (as above) the number of 
emitted neutrinos, the distance to the supernova is $D$, the number of target atoms 
in a given detector is $N_{tar}$, and the neutrino-energy-dependent cross section 
for a given interaction channel is $\sigma{\left(E_{\nu} \right)}$. 
The efficacy of neutrino detection also depends upon the efficiency of the detectors, $\epsilon(E)$, 
where $E$ is the energy of the final-state product (such as an electron). This is approximately 1.0 above some 
detector-dependent threshold and 0.0 below it.  In water Cherenkov detectors such as Super-K
this threshold is $\sim$5 MeV.  Given the detectors we have chosen to highlight, we 
focus on neutrino interactions in water, ice, scintillator, and liquid argon.
SNOwGLoBES imposes a particular energy binning.  As a result, our model spectra, which are grouped into 20 
logarithmically-spaced energy bins from 1 MeV to either 300 (for $\nu_e$s) or 100 MeV 
(for $\bar{\nu}_{e}$s and $\nu_x$s) (Radice et al. 2017), must be mapped onto its grid.  We
do this using a straightforward interpolant, which executes one-dimensional monotonic cubic splines, 
across SNOwGLoBES's energy range from 0.5 to 100 MeV. 

When estimating the statistical errors in the time bins selected when binning the 
signals (and these bin widths can be arbitrary), we multiply the height (signal rate) by 
the bin width, ${\Delta}t$, take the square root, and then divide by the bin width. 
Thus, we calculate the event rate error ($\sigma$) for each bin using the simple formula:
\begin{equation}
\sigma {\Delta}t =  \sqrt{\frac{dN_{det}}{dt} {\Delta}t}\, .
\end{equation}
In this way, we obtain a measure of the error in the signal rate for the given bin width.  This, of course,
is a function of distance.  If we choose a wide bin width, the error in the average signal rate 
around that time bin is perforce lower, but the temporal resolution would be correspondingly diminished.  
As long as the detectors have good time tagging, this procedure is, of course, arbitrary and to taste.
We explore some of the general conclusions using this approach in \S\ref{accuracy}.

\section{Neutrino Oscillations}
\label{osc}

The incoming neutrinos, by their nature, undergo neutrino oscillations.  To incorporate such 
oscillations, we employ the approximate approach of Dighe \& Smirnov (2000) to account for the adiabatic
Mikheyev-Smirnov-Wolfenstein effect in the stellar envelope.  This approach for handling neutrino
oscillation effects in the context of supernova neutrino detection is similar to, or more 
sophisticated than, for example that found in Nakamura et al. (2016), Nikrant et al. (2018), Kawogoe et al. (2010), 
Scholberg  (2018), Tambora et al.  (2014), Abe et al. (2016), Serpico et al. (2012), M\"uller \& Janka (2014),
and Gallo Rosso et al. (2018).  

Our only major assumption is adiabaticity, and we ignore
for the purposes of clarity additional matter oscillation effects in 
the Earth.   Since the direction from which a neutrino burst would come will be from any angle
in an Earth-centered coordinate system, the Earth effects are unpredictable. The
associated Earth column tranversed by the neutrinos is a priori unknown.  Addressing this
a priori, therefore, would add an extra degree of complexity generally ignored in the
literature.  However, slight flavor regeneration while traversing the Earth has less of an effect on the total 
signal rate upon which we focus in this paper than upon the detailed measured spectra.  
With a high neutrino energy resolution (perhaps beyond the capability of the current Super-K, but 
within that anticipated for JUNO), this effect might be discerned and diagnostic. 
The reader is referred to Liao (2016) for a discussion of this issue.
The only non-adiabatic effect of interest may be the encountering 
by the emergent neutrinos of any steep density jumps in the outer
star, perhaps associated with the supernova shock itself (Mirizzi et al. 2016).  However, addressing this
possibility would introduce, we feel, unnecessary and very uncertain complications
associated with special details of the phenomenon and variable outer progenitor structures 
and is beyond the scope of our thesis. Note that neutrino-neutrino self-refraction
effects (Pantaleone 1992; Duan, Fuller, \& Qian 2006) do not obtain for these models, 
due to the dominance of electron lepton number over the neutrino lepton number (Dasgupta, O'Connor, \& Ott 2012; Sarikas et al. 2012).  
It is only for very low-mass massive stars (such as the Nomoto and Hashimoto (1984) 8.8 M$_{\odot}$ model) that 
self-refraction effects might be of interest, but these models are not in our set.

The resulting mappings are:
\begin{equation}
F_{\nu_e} = p F^{0}_{\nu_e} +
\left(1 - p \right) F^{0}_{\nu_x}\, , \\ \nonumber
 F_{\overline{\nu}_e}=  \overline{p} F^{0}_{\overline{\nu}_e} +
\left(1 - \overline{p} \right) F^{0}_{\nu_x} \, ,
\end{equation}
and 
\begin{equation}
4 F_{\nu_x} =
\left(1 - p \right) F^{0}_{\nu_e} +
\left(1 - \overline{p} \right) F^{0}_{\overline{\nu}_e} +
\left(2 + p  + \overline{p} \right) F^{0}_{\nu_x}\, .
\end{equation}
In these equations, the initial flux of a neutrino species is $F^0_i$ and the survival probabilities 
are $p$ and $\overline{p}$. Dighe \& Smirnov (2000) also assume that the total flux of the heavy lepton 
species satisfies:
\begin{equation}
4 F_{\nu_{x}} = F_{\nu_{\mu}} + F_{\nu_{\tau}} +   F_{\overline{\nu}_{\mu}} + F_{\overline{\nu}_{\tau}}\, .
\end{equation}

Utilizing the survival probabilities of Kato et al. (2017) that
incorporate the $ \theta_{13}$ mixing angle, we have for the Normal Hierarchy (NH):
\begin{equation}
p = \sin^2 \theta_{13}\,\,\,\,\,\,\,\, {\rm and}\,\,\,\,\,\,\,\, \nonumber
\overline{p} = \cos^2 \theta_{12} \cos^2 \theta_{13}
\end{equation}
and for the Inverted Hierarchy (IH):
\begin{equation}
p = \sin^2 \theta_{12} \cos^2 \theta_{13}\,\,\,\,\,\,\,\, {\rm and}\,\,\,\,\,\,\,\, \nonumber
\overline{p} = \sin^2 \theta_{13}\, .
\end{equation}
Using these survival probabilities, we, thus, obtain for the Normal Hierarchy:
\begin{equation}
F_{\nu_e} = \sin^2 \theta_{13} F^{0}_{\nu_e} +
\left(1 - \sin^2 \theta_{13} \right)
F^{0}_{\nu_x}\, ,
\label{first_osc}
\end{equation}
\begin{equation}
 F_{\overline{\nu}_e}= \cos^2 \theta_{12} \cos^2 \theta_{13} F^{0}_{\overline{\nu}_e} +
 \left(1 - \cos^2 \theta_{12} \cos^2 \theta_{13} \right)
 F^{0}_{\nu_x}\, ,
\end{equation}
and
\begin{eqnarray}
4 F_{\nu_x} = \left(1 - \sin^2 \theta_{13}\right)
F^{0}_{\nu_e} +
\left( 1 - \cos^2 \theta_{12} \cos^2\theta_{13} \right)
F^{0}_{\overline{\nu}_e} +\\ \nonumber
\left( 2 + \sin^2 \theta_{13} + \cos^2 \theta_{12} \cos^2 \theta_{13} \right)
F^{0}_{\nu_x}
\end{eqnarray}
and for the Inverted Hierarchy:
\begin{equation}
F_{\nu_e} = \sin^2 \theta_{12} \cos^2 \theta_{13} F^{0}_{\nu_e} +
\left(1 - \sin^2 \theta_{12} \cos^2 \theta_{13} \right)
 F^{0}_{\nu_x}\, ,
\end{equation}
\begin{equation}
F_{\overline{\nu}_e} = \sin^2 \theta_{13} F^{0}_{\overline{\nu}_e} +
\left(1 - \sin^2 \theta_{13} \right)
 F^{0}_{\nu_x}\, ,
\label{last_osc}
\end{equation}
and
\begin{eqnarray}
4 F_{\nu_x} =
\left( 1 - \sin^2 \theta_{12} \cos^2 \theta_{13} \right)
F^{0}_{\nu_e} +
\left(1 - \sin^2 \theta_{13} \right)
F^{0}_{\overline{\nu}_e} + \\ \nonumber
\left( 2 + \sin^2 \theta_{12} \cos^2 \theta_{13} + \sin^2 \theta_{13} \right)
F^{0}_{\nu_x}
\end{eqnarray}
When accounting for oscillations, we use the best fit mass-mixing parameters 
from Capozzi et al. (2017): $\sin^2(\theta_{12}) = 2.97\times 10^{-1}$     
and $\sin^2(\theta_{13}) = 2.15\times 10^{-2}$ for both hierarchies.
The high electron densities around the inner core for these models allows us
to ignore self-refraction effects on the emergent neutrino spectra, and, as implied,
we ignore any possible non-adiabatic oscillation effects due to shocks and 
at steep compositional interfaces.

We see from equations (\ref{first_osc}) through (\ref{last_osc}) and the values of the oscillation angles that for 
the normal hierarchy $\nu_e$s in a detector come mostly from $\nu_x$s at the source
and $\bar{\nu}_e$s at the detector are a mix of $\nu_x$s and $\bar{\nu}_e$s at the source.
For the inverted hierarchy, to a slightly greater degree than for the normal hierarchy
$\nu_x$s source $\nu_e$s at the detector, while $\bar{\nu}_e$s in a detector were once 
$\nu_x$s (even more so than for the normal hierarchy).  These mappings are relevant for the
differences in the event rates for the various detectors (differentially sensitive as they are 
to the various neutrino flavors) and hierarchies. 

\section{Underground Detector Characteristics}
\label{detectors}

\subsection{Super-Kamiokande} 
\label{subsec:water}

Super-Kamiokande (Super-K; SK) is a $\sim$50-ktonne water Cherenkov neutrino 
detector in the Kamioka Mines in Japan, arrayed with photomultiplier tubes ($PMTs$), 
configured with a burst monitor for supernova neutrino bursts, and utilizing 
efficient false trigger rejection. It has a fiducial mass/``volume" of 
32.5 ktonnes of ultra-pure water (Abe et al. 2016) and is, therefore, proton-rich.
As such, it is most sensitive (Beck et al. 2013) to electron antineutrinos ($\bar{\nu}_e$) 
via super-allowed charged-current (CC) inverse beta decay (IBD) reaction:
\begin{equation}
\overline{\nu}_{e} + p \rightarrow e^{+} + n\, ,
\end{equation}
which also boasts a low neutrino energy threshold of 1.8 MeV.  Electron neutrinos 
and antineutrinos also interact via the charged-current absorption processes on Oxygen:
\begin{equation}
\nu_{e} + {^{16}O} \rightarrow  e^{-} + {^{16}F}
\end{equation}
and
\begin{equation}
\overline{\nu}_{e} + {^{16}O} \rightarrow  e^{+} + {^{16}N.}
\end{equation}

\begin{figure}
\includegraphics[width=1.0\textwidth, angle = 0]{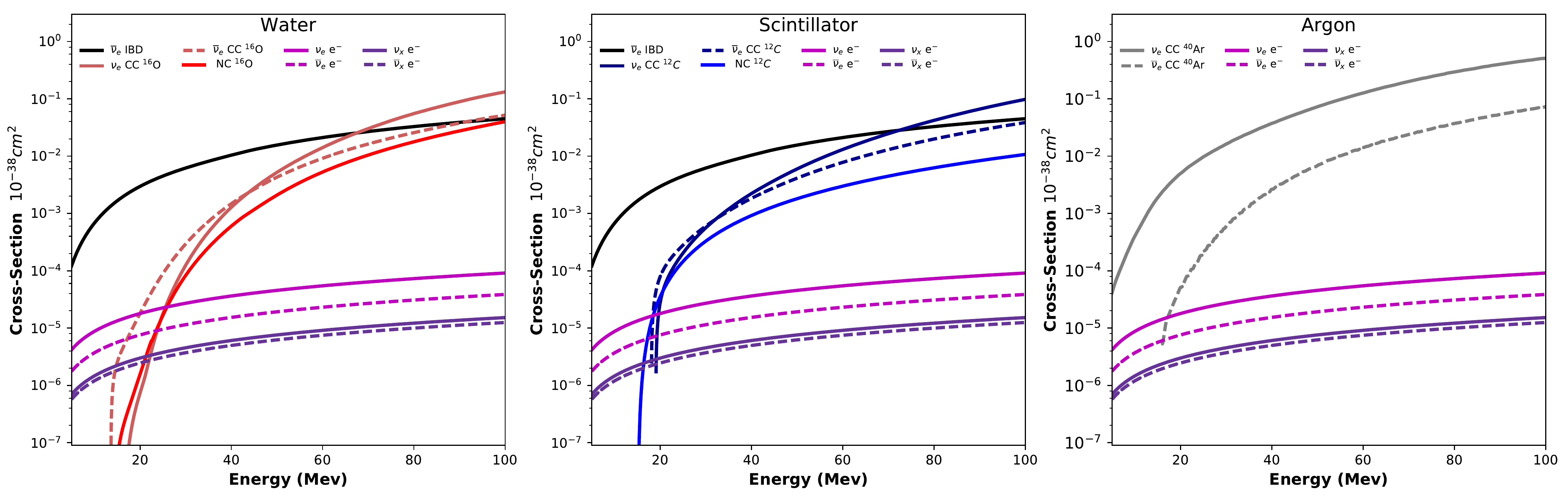}
\caption{Above are the energy-dependent cross sections for the neutrino-matter interactions in water (left),
scintillator (center), and liquid argon (right). These cross sections were provided by the SNOwGLoBES software (Beck et al
2013), and we focus on the neutrino energies up to $\sim$100 MeV.
\label{cross}}
\end{figure}

The final state of CC interactions can also include neutrons and 
deexcitation gammas, and these are detectable.  Detection of these
secondaries can in principle be used to identify the interaction channel.
Figure \ref{cross} depicts the suite of relevant cross sections
for the detectors highlighted in this study, and, as indicated in the leftmost panel, 
that for the IBD dwarfs those for the CC interactions on oxygen.  
All neutrino species interact via the neutral current( NC), and, thus, 
contribute to the signal via the measurment of a deexcitation gamma that accompanies 
the neutral-curent excitation of resident nuclei in reactions such as: 
\begin{equation}
\nu_i + {^{16}O} \rightarrow  \nu_i + {^{16}O^*}\, .
\end{equation}
All neutrino species scatter off electrons via the process:
\begin{equation}
\nu_i + e^{-} \rightarrow  \nu_i + e^{-}\, ,
\end{equation}
but this process is subdominant.
however, in this Compton-like process, the directionality of the incident neutrino is partially
preserved, and this fact allows neutrino-electron scattering to be the dominant means, 
using neutrinos alone, to determine the direction of the supernova. 
As seen in Figure \ref{cross}, the $\nu_e - e^-$ scattering cross section exceeds that 
for $\bar{\nu}_e - e^-$ scattering, and these are followed by those for  $\nu_{x} - e^-$ scattering.
In order to enhance its senstivity to the IBD reaction and the assocaited secondary neutron, 
Super-K (and the follow-on Hyper-K, at $\sim$250 ktonnes) might in the future be spiked 
with gadolinium (Beacom \& Vagins 2004; Laha \& Beacom 2014).

\subsection{IceCube} 
\label{subsec:ice}
IceCube is currently the largest neutrino observatory, with an effective mass 
of $\sim$3.5 Mtonnes of pure water ice (Abbasi et al. 2011).  Thus comprised of
water, the suite of relevant neutrino-matter interactions is the same as
in water Cherenkov detectors such as Super-K (\S\ref{subsec:water}).  However, given its high 
energy threshold (above $\sim$100 MeV), it is not currently capable of identifying
individual supernova neutrino interactions, nor the energies of final-state electrons,
gammas, or neutrons. Rather, the interaction of supernova neutrinos, predominantly by 
the IBD, will register in IceCube as a sudden increase in the background noise count 
rate.  There are 5160 modules, each of which would detect the neutrino (mostly 
$\bar{\nu}_e$s via the IBD) by the measurement in the phototubes of each module 
of the Cherenkov photons created by the neutrino's charged secondaries.  The 
Cherenkov photon yield is $\sim$178$\varepsilon_e$, where $\varepsilon_e$ is 
the energy of the secondary positron or electron in MeV.   This yields an effective 
volume per detector for the measurement of a single photon (on average and to compete with the 
noise in the module) of $\sim$590 cubic meters at $\varepsilon_e \sim 20$ MeV (K\"opke et al. 2011).
Multiplying this number by 5160 yields an equivalent detector mass of $\sim$2.8 Mtonnes.
Since the sensitivity and signal yield of a supernova in IceCube depend upon numerous 
detector-specific systematics, such as angular sensitivity and absorption length,
a Monte Carlo analysis of the detector capabilities, best performed by the IceCube team, 
is necessary.  Therefore, for our purposes we will assume that the effective mass for 
100\% detection is 3.5 Mtonnes of water ice and the reader is encouraged to scale 
our theoretical signal rates to any updated effective mass.    

For a galactic supernova, the sharp increase in the background
rate over the short few-second time period associated with a supernova 
neutrino burst will be unmistakable.  The background rate 
per detector module with a 0.25-millisecond deadtime setting is estimated to be 
$\sim$286 Hertz (K\"opke et al. 2011)\footnote{This deadtime approach, along
with the sampling rate that yielded a temporal resolution of $\sim$2 milliseconds, 
may recently have been improved.  Such changes will affect the per-module noise rate. 
However, for specificity and for the purposes of this study, we use the older detector 
characterisitics as envisioned in Abbasi et al. (2011) and K\"opke et al. (2011).}.  
It is assumed that the noise in these modules is uncorrelated and 
Poissonian.  This yields a sigma for the collective background fluctuation
of $\sim$1200$\Delta{t}$, where $\Delta{t}$ is the width of a time bin in seconds.
This, together with the Poissonian fluctuation in the signal itself, sets the total noise
floor (added in quadrature) against which to compare the average supernova signal to 
determine detectability.  Since the Poissonian variation in the supernova signal goes
as the square root of the signal, and, hence, inverse-linearly with distance, while the IceCube 
detector background is independent of distance, the overall noise level is dominated by 
the signal fluctuations at smaller distances (where the signal-to-noise will be large) 
and by the detector noise at large distances.  This is not the case for the other 
detectors, for which there is effectively no background during a $\sim$second-long 
integration.  

Nevertheless, IceCube, given its large effective mass, will be an excellent means 
to measure the neutrino- and neutrino-energy-integrated light curve. Therefore, 
though IceCube cannot provide event-by-event information, particle characterization, 
directionality, nor energy, it is still an exceptional detector for
measuring the temporal development of the total neutrino flux.  This can help constrain 
structures in the time signal such as accretion and cooling, and may provide good
progenitor discrimination (see \S\ref{ranges} and \S\ref{accuracy}).

\subsection{DUNE} 
\label{subsec:argon}

The Deep Underground Neutrino Experiment (DUNE) is a $\sim$40-ktonne 
Liquid Argon Time Projection Chamber (LAr TPC) that is being built at 
the Sanford Laboratory in South Dakota's Homestake Mine (Acciarri et al. 2016). 
This class of detector has high sensitivity to electron neutrinos via the charged-current reaction:
\begin{equation}
\nu_e + {^{40}Ar} \rightarrow  e^{-} + {^{40}K^*}\, ,
\end{equation}
which has a large per-nucleus cross section (see Figure \ref{cross}) and a neutrino 
energy threshold of only 1.5 MeV (Botella 2016). DUNE will provide the best 
electron-neutrino light curve, and in fact will be the only detector to provide a high-statistics 
$\nu_e$ signal (Ankowski et al. 2016). Electron antineutrinos also interact with Argon 
via the charged current reaction:
\begin{equation}
\overline{\nu}_e + {^{40}Ar} \rightarrow  e^{+} + {^{40}Cl^*}\, .
\end{equation}
This interaction's cross-section, as seen in Figure \ref{cross}, is as much as two orders of 
magnitude smaller than that of its electron-neutrino counterpart. Of course, all species of neutrinos 
will contribute to the signal by elastically scattering off electrons. We do not in this study include
neutral-current interactions in DUNE;  there are still ongoing studies to determine the neutral-current 
interaction rates Argon, but these are currently unavailable in the SNOwGLoBES software 
(Kemp 2017; Beck et al. 2013). In a liquid argon detector, the final-state charged particles 
deposit kinetic energy along an ionization trail in the liquid argon. In a TPC, a voltage 
is applied across the liquid argon to cause charges to drift towards the anode to be collected 
on wire planes.  The times of arrival to the wire plane allow the three-dimensional tracks to be 
reconstructed. Individual articles can be identified by the rate of energy loss (Ankowski et al. 2016)
and, therefore, DUNE is capable of resolving the energies of the final-state charged particles,
from which the incident neutrino energy can be estimated. 
Without a tagging mechanism, the electrons from CC interactions cannot be distinguished from 
the elastically-scattered electrons, but the CC interaction on Argon has such a large cross section
that it dominates the signal. In summary, DUNE will provide the most accurate estimates of the
electron neutrino emissions of a core-collapse supernova.

\subsection{JUNO} 
\label{subsec:scintillator}

The Jiagmen Underground Neutrino Observatory is a Liquid Scintillator Antineutrino Detector 
currently under construction in Southern China. This neutrino detector will utilize 
$\sim$20 kilotons of linear alkyl-benzene hydrocarbon scintillator, and employ 
17,000 high-quantum-efficiency photomultiplier tubes (Grassi 2016). Its primary research 
objective is to study neutrino oscillations utilizing reactor-produced electron 
antineutrinos. However, like water-cherenkov detectors, scintillator is proton-rich
and is, therefore, sensitive to electron antineutrinos and the high-statistics IBD signal. 
In JUNO, scintillation light is emitted as charge particles as they lose energy in the 
medium, and that light is captured by PMT's. In additon to the IBD, charge-current absorption
on carbon via reactions such as:
\begin{equation}
\nu_e + {^{12}C} \rightarrow  e^{-} + {^{12}N}
\end{equation}
\begin{equation}
\overline{\nu}_e + {^{12}C} \rightarrow  e^+ + {^{12}B}.
\end{equation}
contribute to the signal in the detector.

The charged-current interactions for the electron and anti-electron neutrinos have thresholds 
of 17.34 MeV and 14.39 MeV, respectively (An et al. 2015).  As indicated in Figure \ref{cross}, at
20 MeV the IBD cross section is $\sim$40 times larger than that for the $\overline{\nu}_e-{^{12}C}$ 
CC interaction. As in other detectors, all neutrino species elastically scatter off of electrons. 
JUNO is capable of distinguishing the IBD reaction because its final-state neutrons will 
quickly capture on a proton and emit a 2.2 MeV gamma ray. Pulse-shape discrimination is used 
to distinguish electrons from positrons, thus allowing CC absorption and electron scattering 
to be distinguished (Franco et al. 2011).  Neutral-current interactions with carbon:
\begin{equation}
\nu_i + {^{12}C} \rightarrow  \nu_i + {^{12}C^*},
\end{equation}
involving all species are possible, and these produce observable deexcitation gammas, 
but JUNO will have limited senstivity to them. However, JUNO should have good event-by-event 
energy resolution, though directional information
will be difficult to acquire in the scintillator.

\section{Supernova Models}
\label{snmodels}

Between Radice et al. (2017), Burrows et al. (2018), and Vartanyan et al. (2018), 
we calculated a collection of supernova models starting from the 9, 10, 11, 16, 17, 19, 
and 21 M$_{\odot}$ progenitor models of Sukhbold et al. (2016) and 
integrated out to $\sim$1 second after core bounce. The 1D variants did not explode,
while the 2D variants did. These models, while not the final word, collectively 
capture the wide range of detailed model expectations for the neutrino
emissions and signatures that have emerged from most recent theoretical work on 
the supernova mechanism. The multi-group radiation/hydrodynamic code F{\sc{ornax}} 
(Skinner et al. 2018; Burrows et al. 2018) was used with twenty energy 
groups for each neutrino species\footnote{For these simulations, 
we follow the $\nu_e$, $\bar{\nu}_{e}$, and ``$\nu_{\mu}$" species, where
the four species, $\nu_{\mu}$, $\bar{\nu}_{\mu}$, $\nu_{\tau}$, and $\bar{\nu}_{\tau}$ 
are lumped together into ``$\nu_{\mu}$".  For the $\nu_e$ types, the neutrino
energy $\varepsilon_{\nu}$ varied logarithmically from 1 MeV to 300 MeV, while it varied
from 1 MeV to 100 MeV for the $\bar{\nu}_{e}$s and $\nu_{\mu}$s.}. The radial coordinate, $r$, ran 
from 0 to 20,000 kilometers (km) in 608 zones (both 1D and 2D), while the polar angular grid 
spacing (in 2D) covered the full $180^\circ$ and varied smoothly in 256 zones from
$\approx 0.95^\circ$ at the poles to $\approx 0.65^\circ$ at the equator. 
We used the LS220 (Lattimer \& Swesty 1991) nuclear equation of state 
(EOS) for the 9, 10, and 11 M$_{\odot}$ models and the SFHo 
nuclear equation of state (Steiner et al. 2013) for the rest,
the approximate general relativistic potential formalism of Marek et al. 
(2006), and GR redshifts were included in the transport.  The many-body 
correction of Horowitz et al. (2017) was incorporated into the full
collection of neutrino-matter interaction rates (Burrows, Reddy, \& Thompson 2006)
and inelastic scattering off electrons and nucleons as described in Burrows et al. (2018)
was included. In the 2D simulations, seed perturbations were include in only the
10-M$_{\odot}$ model (Radice et al. 2017), while numerical noise was allowed to 
seed the turbulence in the other 2D models. The products used to determine 
signals in underground detectors were the associated emergent luminosity 
spectra for each neutrino species as a function of time.

Quite generally, 1D and 2D emissions are similar
until explosion. As previously stated, after explosion they differ 
due to the cessation of mass accretion onto the core at explosion, 
which is a major source of neutrino power.  Qualitatively, the two sources of
neutrino emission are diffusion from the core and accretion power.  It is this transition to the phase of
emission without accretion that distinguishes the exploding models, represented in this paper
by the 2D models.  Even if core collapse does not yield a supernova, and even if there 
is no optical counterpart, the neutrino signal from such a ``failed" core will easily 
be observable throughout our galactic neighborhood in the collection of neutrino 
facilities we highlight in this paper.  It is the differences between non-exploding models
and exploding models (represented in this paper by 1D versus 2D models), due to
the cessation of accretion in the latter, that we contend can be distinguished when
the next neighborhood core collapse erupts.

As a copious literature shows (see Burrows 2013, and references 
therein), multi-D is crucial to explosion due to the turbulence behind the shock wave, turbulence that
can be manifest only in 2D or 3D. However, until explosion and the cessation of accretion,
this turbulence has only modest effects on the overall neutrino emissions to infinity.  Hence,
before explosion, the neutrino emissions in 1D, 2D, and 3D simulations for a given progenitor are very
similar.  All models (1D, 2D, 3D) emit roughly like spheres - the multi-D variations due to turbulence
washing out.  The models in the sense of neutrino light curves are all pseudo-spherical.
The role of multi-D is to introduce convection/turbulence to allow explosions by that agency
and 2D and 3D neutrino emissions will be similar.  The only difference expected between 2D and
3D models would be in the time of explosion, and, hence, in the time accretion stops and that
component driving neutrino power, therefore, subsides.  This will introduce only a quantitative
difference between 2D and 3D, i.e. the time of explosion and the time of transition to diffusion
power alone could be slightly different.  In this paper, we are highlighting the qualitative
differences between exploding and non-exploding models to identify a test of the theory
that predicts this.  Hence, the specific time of explosion is secondary to the thesis of our paper
$-$ the existence of the transition is primary.  This has never before been falsified for a supernova
model, making the detection of a neutrino burst and its time structure by the detection
of events in underground laboratories a direct experimental test of supernova theory.  Since
calculating a multitude of 3D models is very expensive (each model can take a month or two
to simulate on supercomputers), our 2D models are reasonable stand-ins for multi-D for the
purposes of this paper.

We note that in our models we don't see any clear signs of a SASI (Blondin, Mezzacappa, 
\& DeMarino 2003) modulation in the neutrino signal rates (Tamborra et al. 2014).  That 
does not mean the SASI is not there, merely that the SASI is sub-dominant and 
obscured by the stronger neutrino-driven convection component.  Specifically, we don't witness 
a near-monochromatic frequency modulation in the neutrino signal seen by 
some others (e.g., Tamborra et al. 2013,2014i; Kuroda et al. 2017).  It may be that SASI-related signal modulations 
are associated with non-axisymmetric spiral modes seen in some non-exploding 3D simulations,
and this possibility deserves further exploration. Such a feature would not be captured by our 
2D models.  However, in Burrows et al. (2012) and Vartanyan et al. (2018a), we argue that the SASI is clearly manifest only when 
the neutrino fluxes are too low to drive explosions, but when they are enough to lead to supernovae there is no
clear tone, nor identifiable SASI feature in the neutrino signal rates. This, again, is what we see here, 
but the reader should note alternate points of view (Lund et al. 2010,2012).

Furthermore, we do not see the LESA phenomenon (Tamborra et al. 2014) in any of our simulations (see 
also Dolence et al. 2015). We speculate (but have not proven) that it is an artefact of the use of the ray-by-ray 
approach (Skinner et al. 2016), and have studied it extensively in both 2D and 3D (Vartanyan et al. 2018b, 
in preparation).  Be that as it may, the LESA would merely result, if it existed, in an 
angular asymmetry in the neutrino lepton flux, and such an angular asymmetry is not 
observable from a single direction (i.e., when obtaining data only at Earth).

\section{Results}
\label{results}

Table \ref{tab:PPer} summarizes the total number of events at 10 kiloparsecs (kpcs)
expected in all detectors during the first $\sim$second after bounce 
for all models (both 1D and 2D) and for the three contexts of no-oscillation 
(NO), the normal hierarchy (NH), and the inverted hierarchy (IH).  
As expected, IceCube 
should experience by far the largest number of events at 10 kpc, of order $10^5$, during this time interval, 
though it can not distinguish interaction channel nor neutrino energy nor type. Nevertheless,
the entire galaxy should be within reach for all models, whether NH or IH obtains.  
The total signals in the other detectors during this earlier dynamical 
phase are still formidable and suggest that the entire galaxy is well within reach by them.  
As expected, among these the more massive facilities (such as Super-K, and later Hyper-K) boast 
the larger event yields.  At the low progenitor mass end, the 1D and 2D models yield roughly similar
integrated event yields (though different light curves in detail).  This is due to the earlier
explosion times post-bounce of such progenitors and to their steeper density profiles.  
The latter translates into a more rapid decrease in the accretion rate and, hence, 
the accretion powered neutrino luminosity, before explosion, so that the accretion-powered phase 
is both weaker and shorter-lived. Diffusion from the core then starts to ascend in relative 
importance.

\begin{figure}
\includegraphics[width=1.00\textwidth, angle = 0]{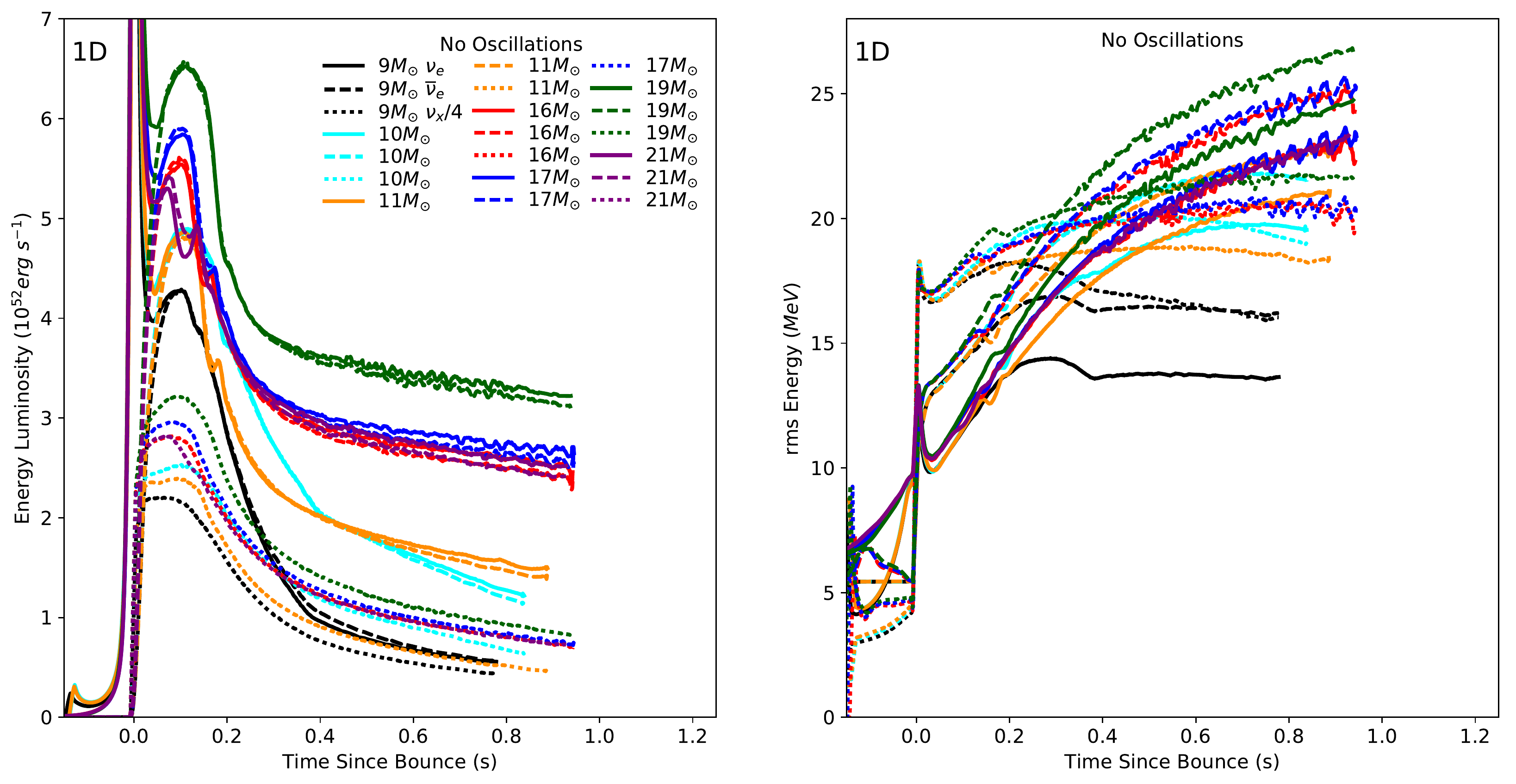}
\includegraphics[width=1.00\textwidth, angle = 0]{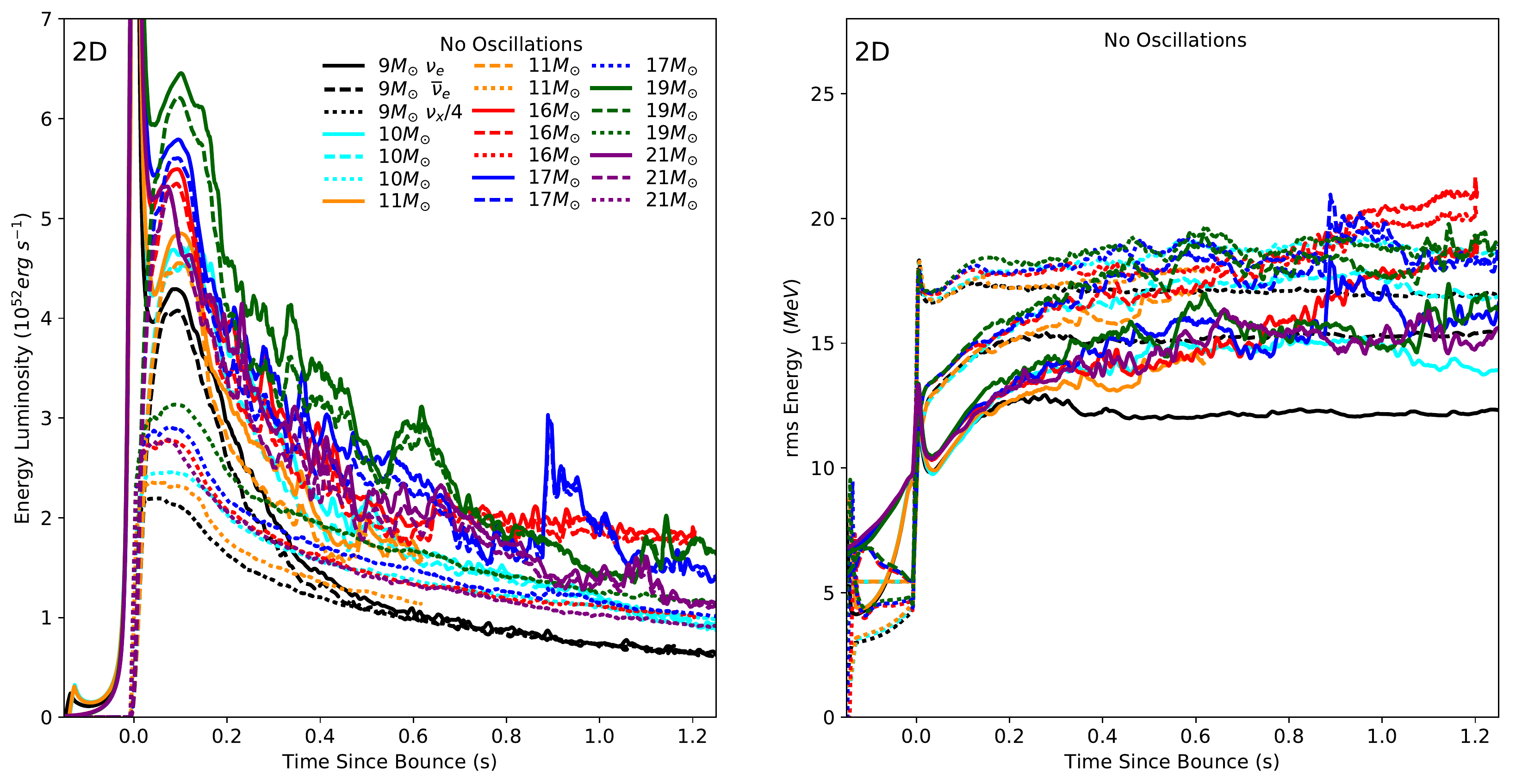}
\caption{Energy luminosities (in units of 10$^{52}$ ergs s$^{-1}$) and root-mean-square (rms) neutrino energies 
(in units of MeV) for all the 1D non-exploding models (top two figures) and 2D exploding models (bottom two 
figures) for the no-oscillation case. The solid curves are for the $\nu_e$s, the dashed curves are for the $\bar{\nu}_e$s,
and the dotted curves are for the ``$\nu_{\mu}$s" ($\times\frac{1}{4}$).  
\label{lum_rms_noosc}}
\end{figure}

\begin{figure}
\includegraphics[width=1.00\textwidth, angle = 0]{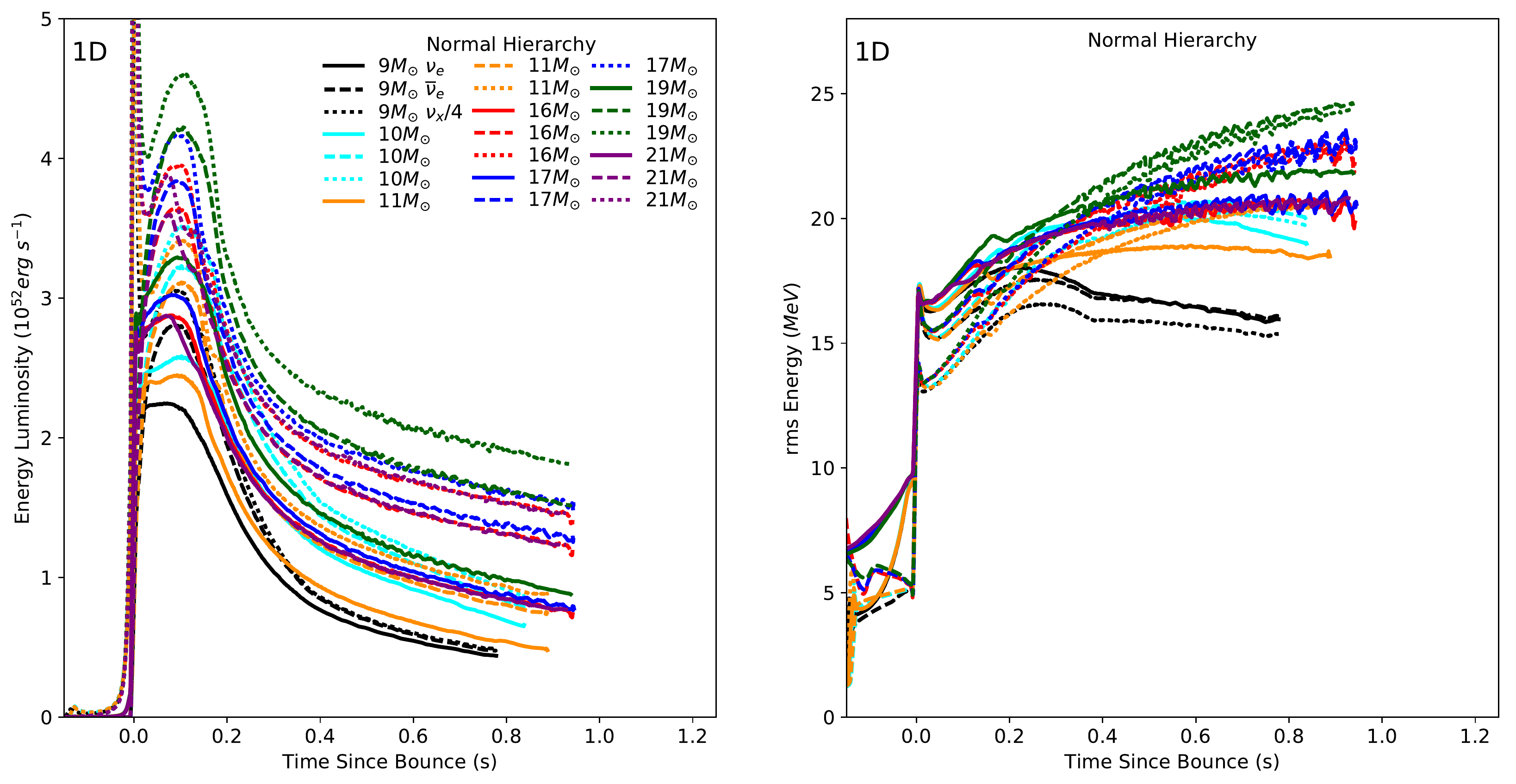}
\includegraphics[width=1.00\textwidth, angle = 0]{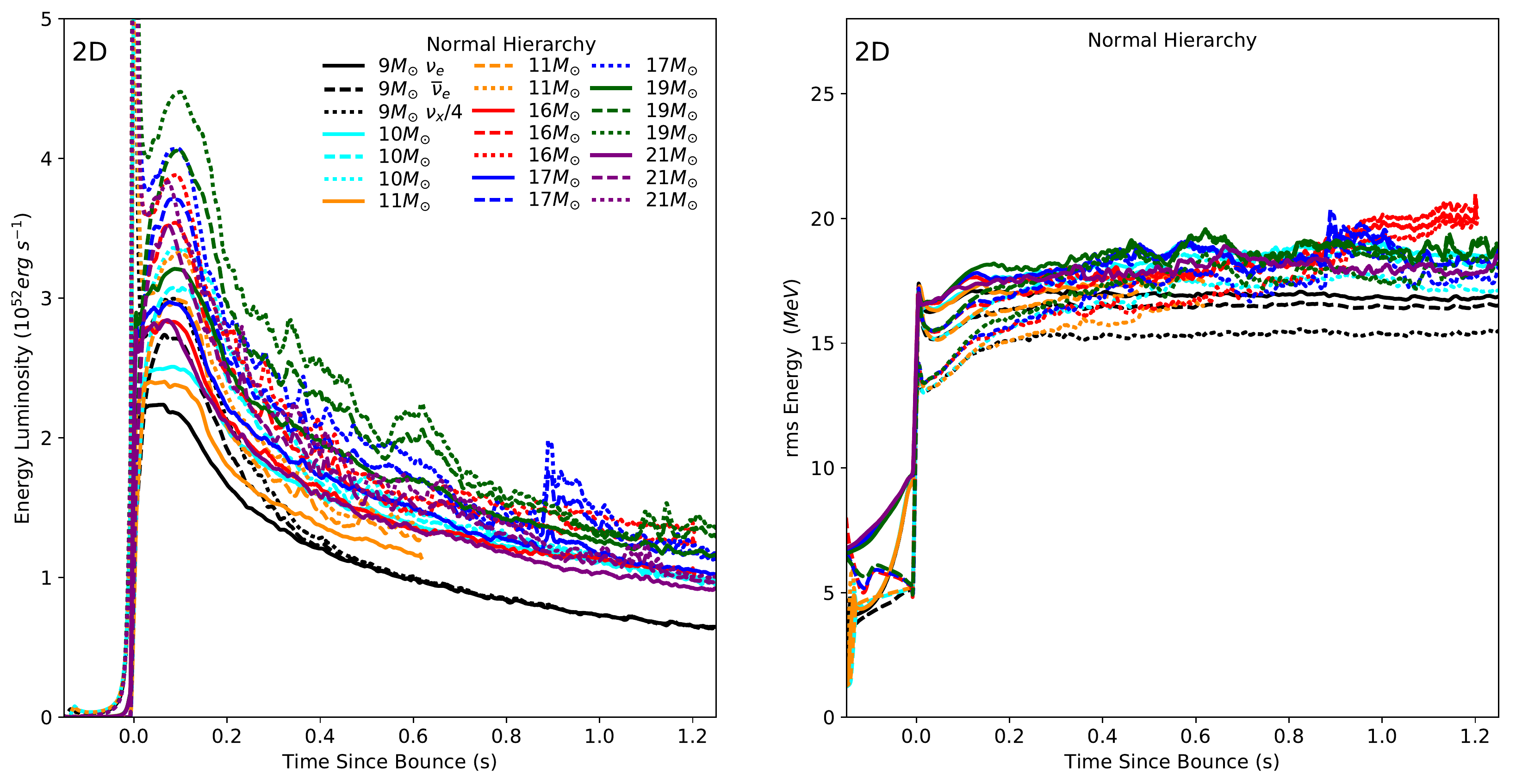}
\caption{Energy luminosities (in units of 10$^{52}$ ergs s$^{-1}$) and root-mean-square (rms) neutrino energies 
(in units of MeV) for all the 1D non-exploding models (top two figures) and 2D exploding models (bottom two 
figures) for the normal-hierarchy oscillation case. The solid curves are for the $\nu_e$s, the dashed curves are for the $\bar{\nu}_e$s,
and the dotted curves are for the ``$\nu_{\mu}$s" ($\times\frac{1}{4}$).
\label{lum_rms_nh}}
\end{figure}

\begin{figure}
\includegraphics[width=1.00\textwidth, angle = 0]{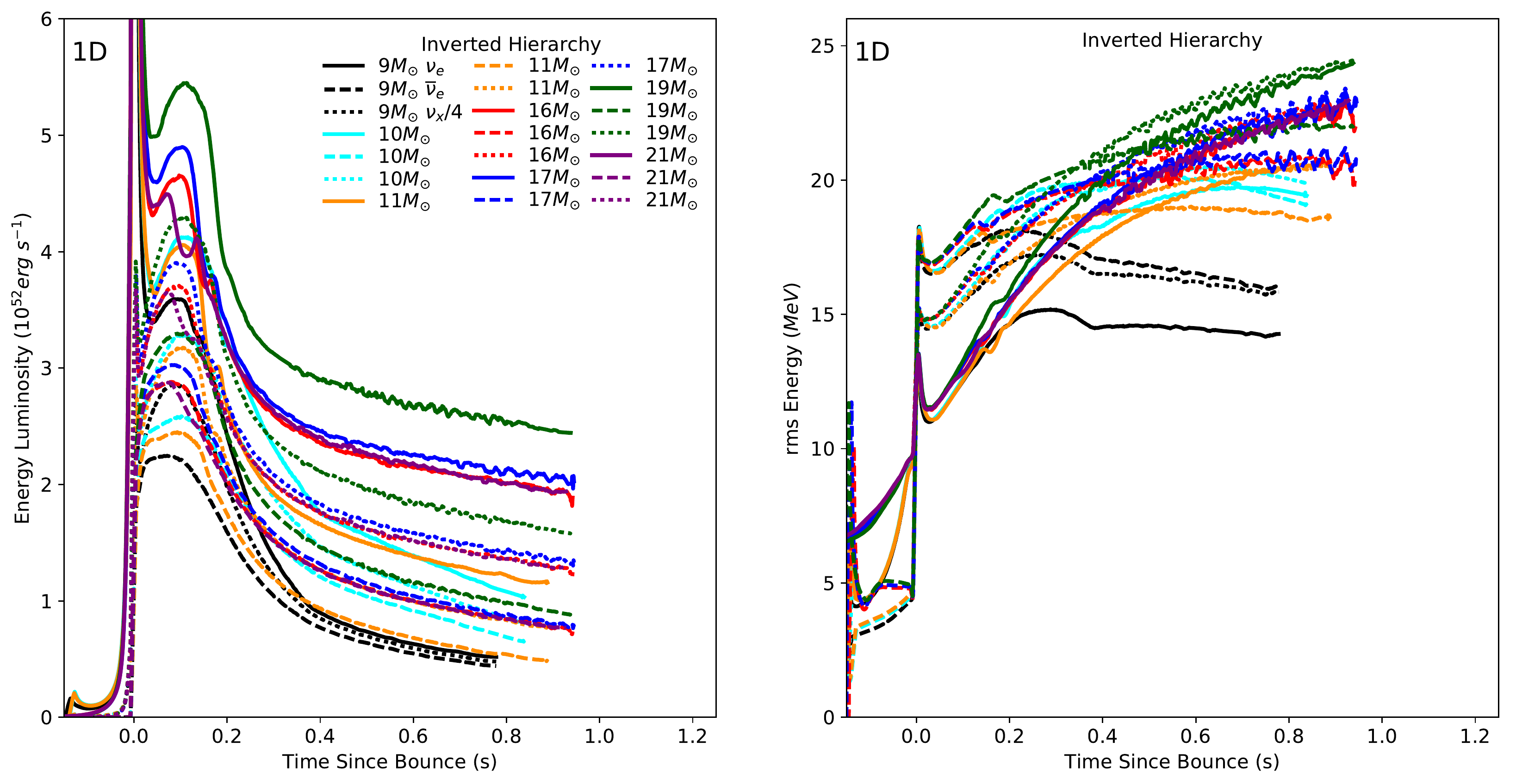}
\includegraphics[width=1.00\textwidth, angle = 0]{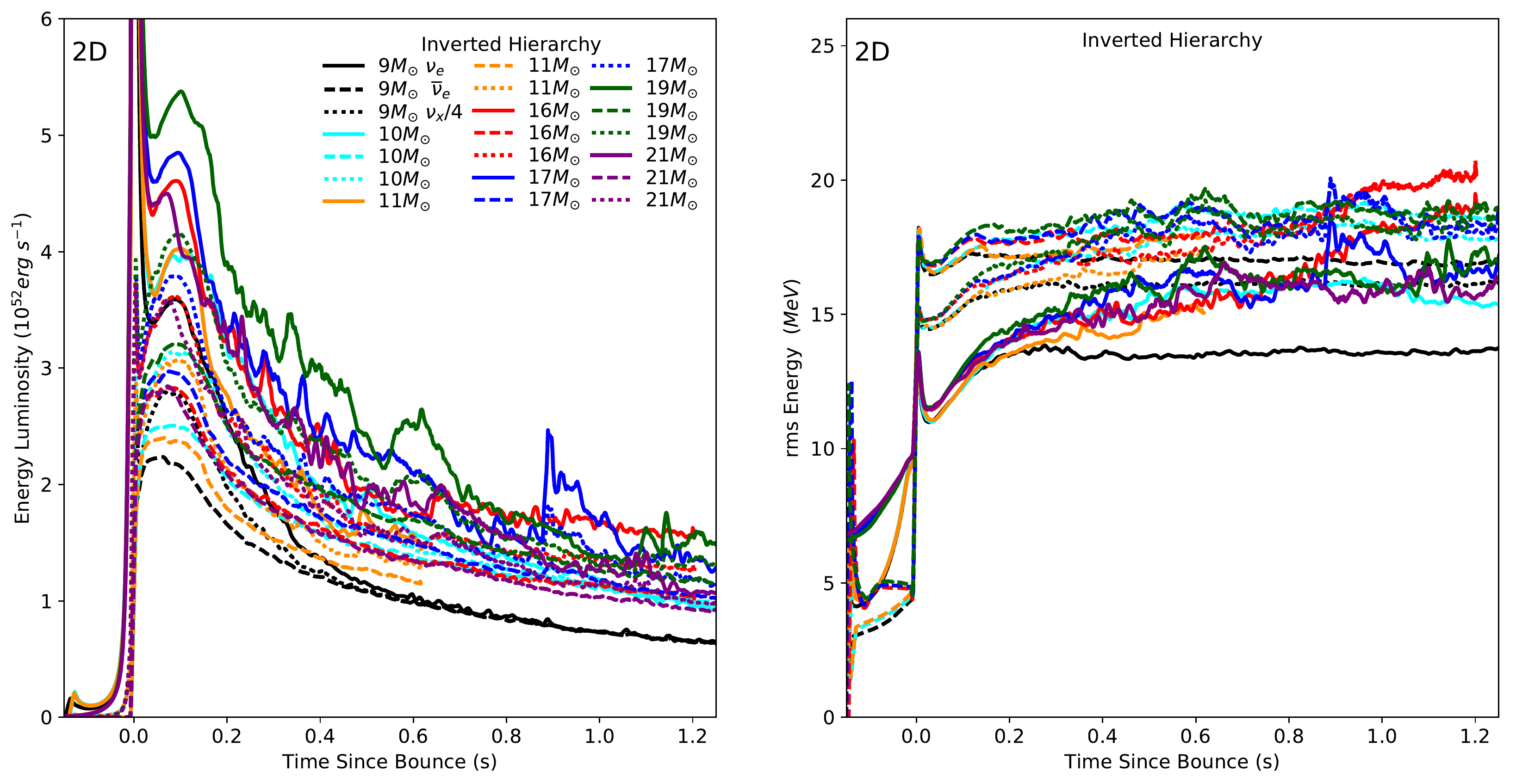}
\caption{Energy luminosities (in units of 10$^{52}$ ergs s$^{-1}$) and root-mean-square (rms) neutrino energies 
(in units of MeV) for all the 1D non-exploding models (top two figures) and 2D exploding models (bottom two 
figures) for the inverted-hierarchy oscillation case. The solid curves are for the $\nu_e$s, the dashed curves are for the $\bar{\nu}_e$s,
and the dotted curves are for the ``$\nu_{\mu}$s" ($\times\frac{1}{4}$).
\label{lum_rms_ih}}
\end{figure}

Figures \ref{lum_rms_noosc}, \ref{lum_rms_nh}, and \ref{lum_rms_ih} provide the energy luminosities
and RMS neutrino energies for all models (both 1D [dashed] and 2D [solid]) and neutrino species
and for the no-oscillation, normal-hierarchy,  and inverted-hierarchy cases.  These curves
depict input quantities before convolution in the various detectors, but after our oscillation
model operates (if it does).  The no-oscillation numbers are what the supernova models give
and the NH and IH numbers reflect the result of applying the oscillation model described
in \S\ref{osc}. For higher-mass progenitors, the corresponding event yields from the 2D 
(exploding) models should be generally lower than those for the 1D models.  In 
these cases, the continuation of significant accretion due to the failure in 1D 
to explode translates into higher neutrino luminosities for longer times.  
In addition, the average and root-mean-square (RMS) neutrino energies of the more massive 
progenitors at the later times are slightly higher than their corresponding 2D counterparts.
The upshot should be significantly higher event rates for the 1D models at later times for the progenitors 
with shallow mass density profiles (high ``compactness"; O'Connor \& Ott 2011,2013).  We suggest that the distinctly different 
event light curves for 1D and 2D, particularly for the higher compactness models
(mostly for higher progenitor mass, with exceptions), reflect in a generic sense 
the difference between non-exploding and exploding models. Factors of more than two 
in the event rates at later times (after hundreds of milliseconds) are possible.
Note that one expects the temporal fluctuations in the event rates and inferred neutrino 
luminosities for the 2D (exploding) models to be larger than in the 1D cases.  This is
a consequence of turbulence in the multi-D models (impossible in 1D) and the associated
episodic nature of plume accretion onto the PNS core before, during, and just after explosion when 
asphericities are allowed.  Fluctuations in the energy lumninosities of as much as 
$\sim$25\% on timescales of $\sim$10 to $\sim$100 milliseconds are seen.  The detection
of such temporal structures could be used to constrain post-bounce and post-explosion accretion
rate variations in angle and time.

\subsection{Neutrino Light Curves}
\label{light}

\begin{figure}
\includegraphics[width=0.95\textwidth, angle = 0]{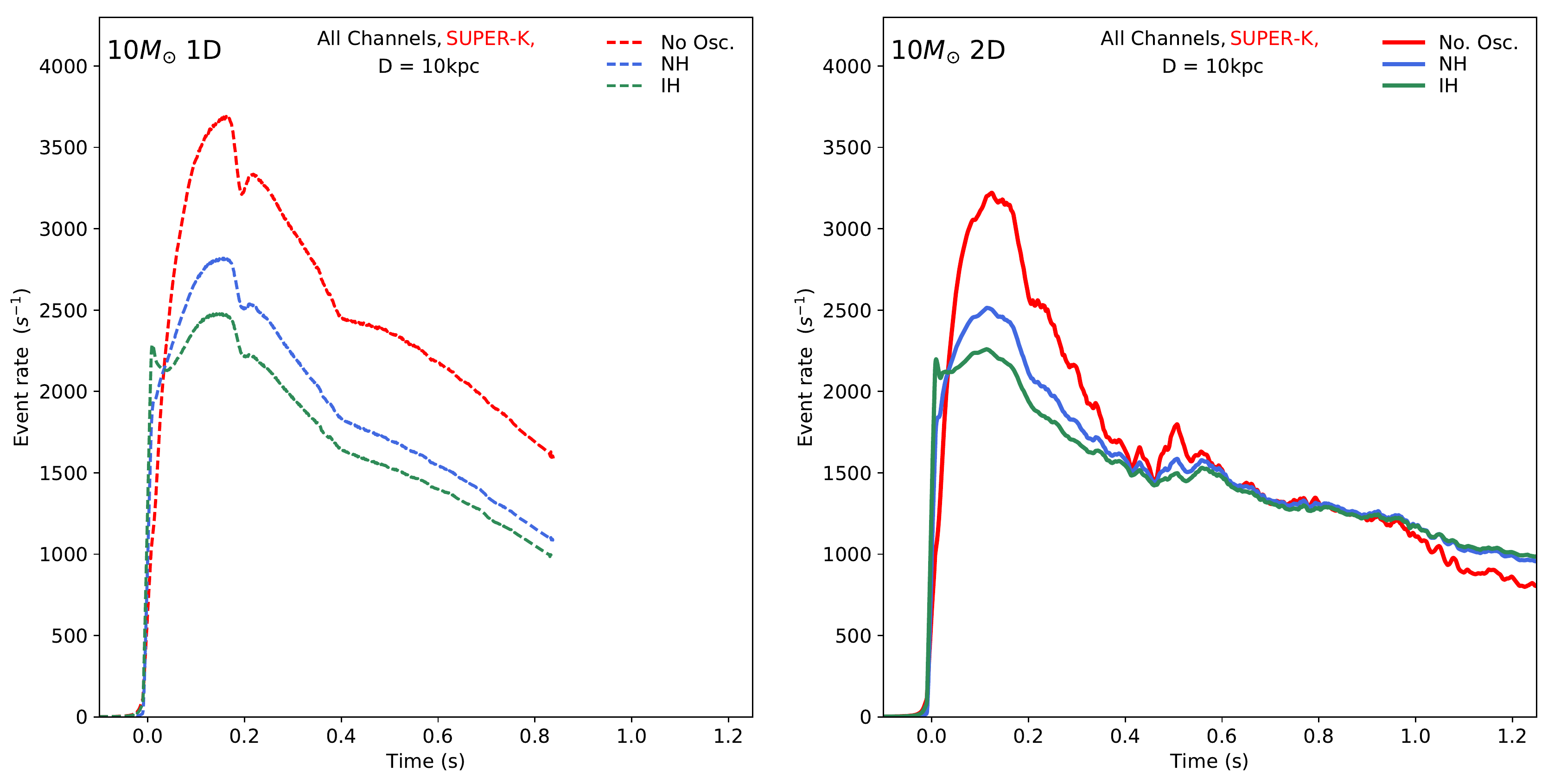}
\includegraphics[width=0.95\textwidth, angle = 0]{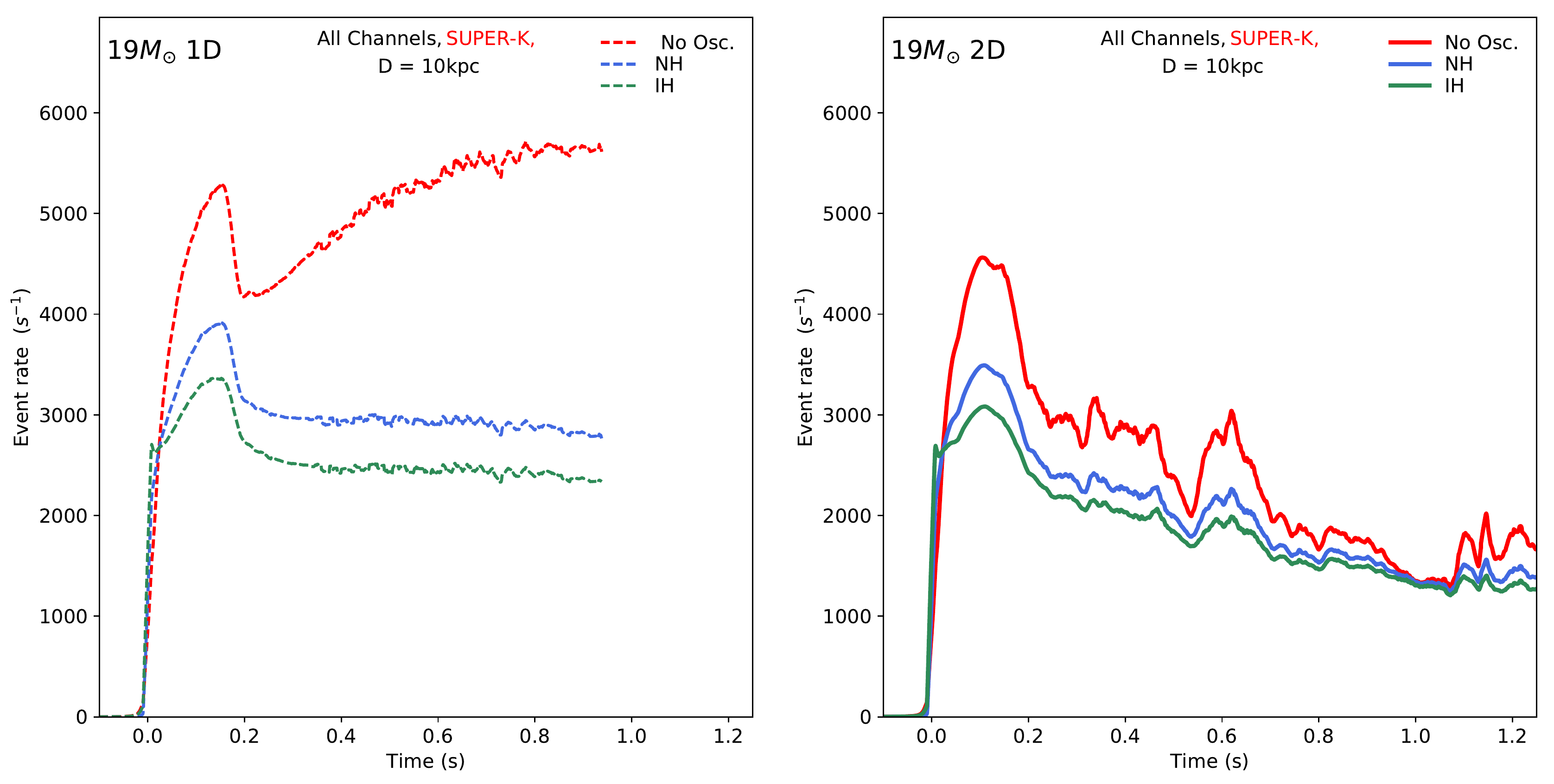}
\caption{The total, all-channel, event rates (in units of s$^{-1}$) in {\bf Super-K} at 10 kiloparsecs 
for all three oscillation models in 1D (left) and 2D (right) for the 10-M$_{\odot}$ progenitor model 
(top two plots) and the 19-M$_{\odot}$ progenitor model (bottom two plots).  
Note that the 1D models are all here plotted as dashed, while the 2D models are all solid.
The no-oscillation case is in red, the normal-hierarchy case is in blue, and the inverted 
hierarchy case is in green.
\label{events_superk_10_19_allosc}}
\end{figure}

\begin{figure}
\includegraphics[width=0.95\textwidth, angle = 0]{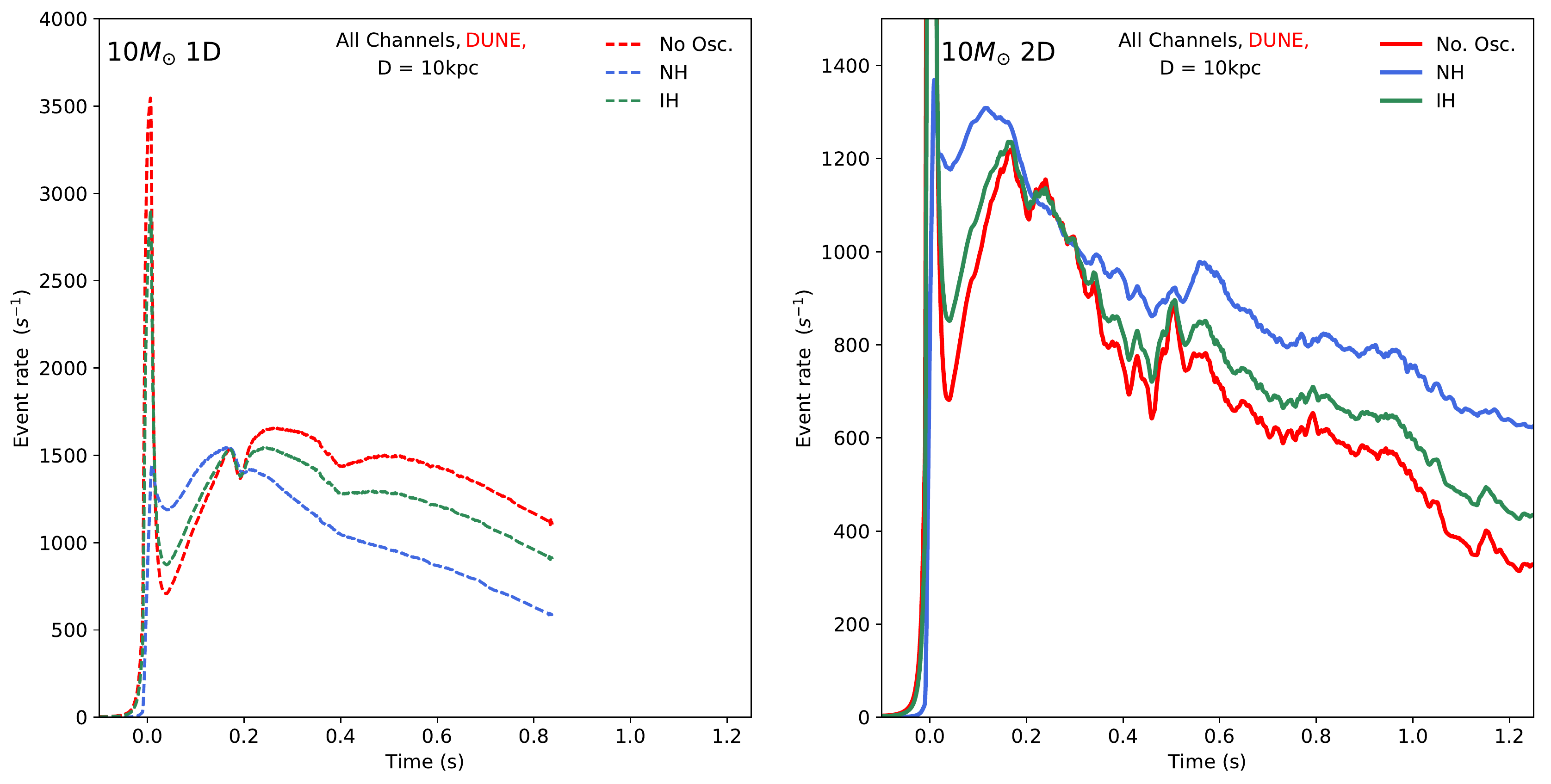}
\includegraphics[width=0.95\textwidth, angle = 0]{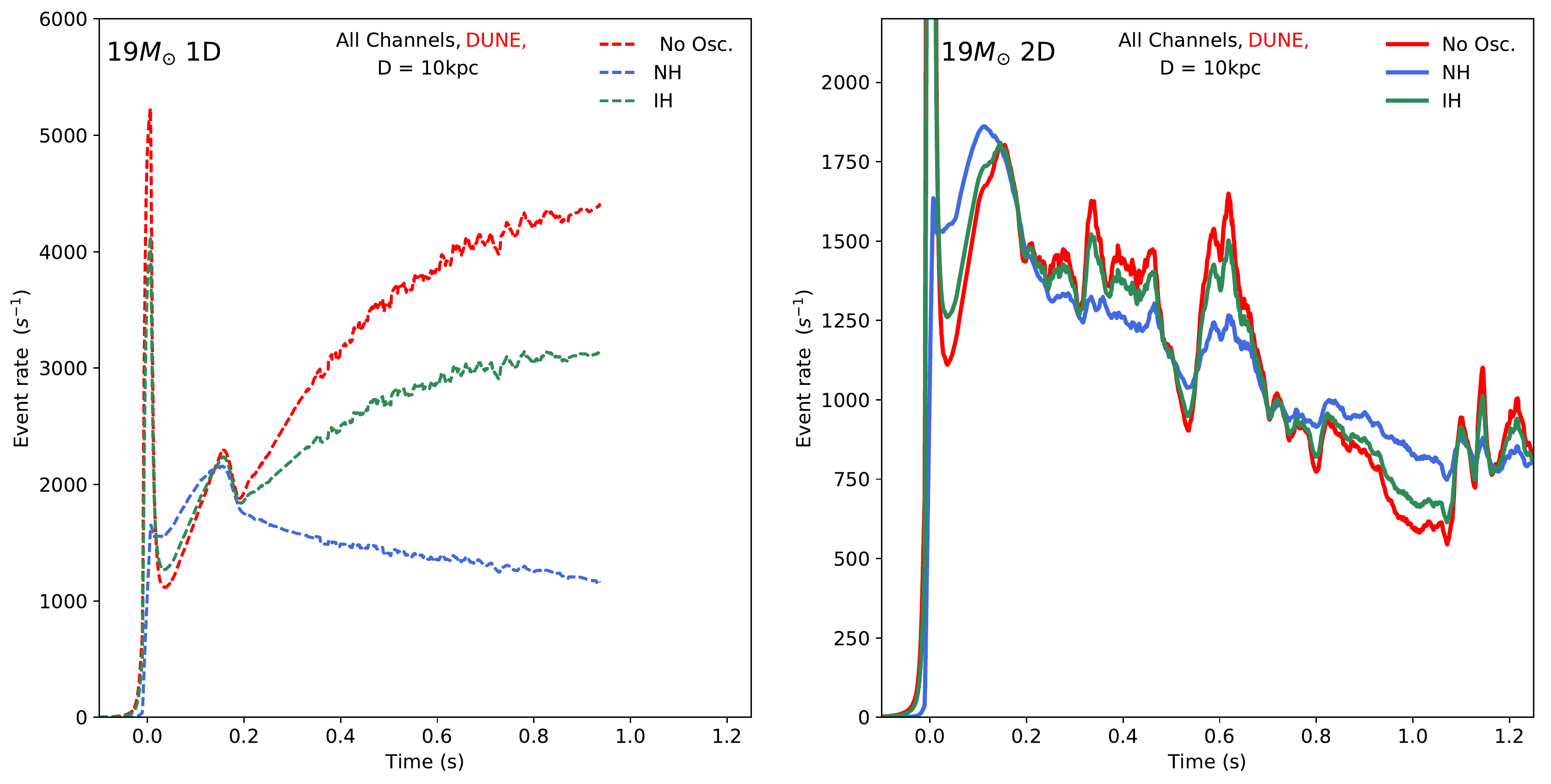}
\caption{The total, all-channel, event rates (in units of s$^{-1}$) in {\bf DUNE} at 10 kiloparsecs 
for all three oscillation models in 1D (left) and 2D (right) for the 10-M$_{\odot}$ progenitor model 
(top two plots) and the 19-M$_{\odot}$ progenitor model (bottom two plots).
As in Figure \ref{events_superk_10_19_allosc}, the 1D models are all plotted as dashed, 
while the 2D models are all solid.
\label{events_dune_10_19_allosc}}
\end{figure}

\begin{figure}
\includegraphics[width=0.95\textwidth, angle = 0]{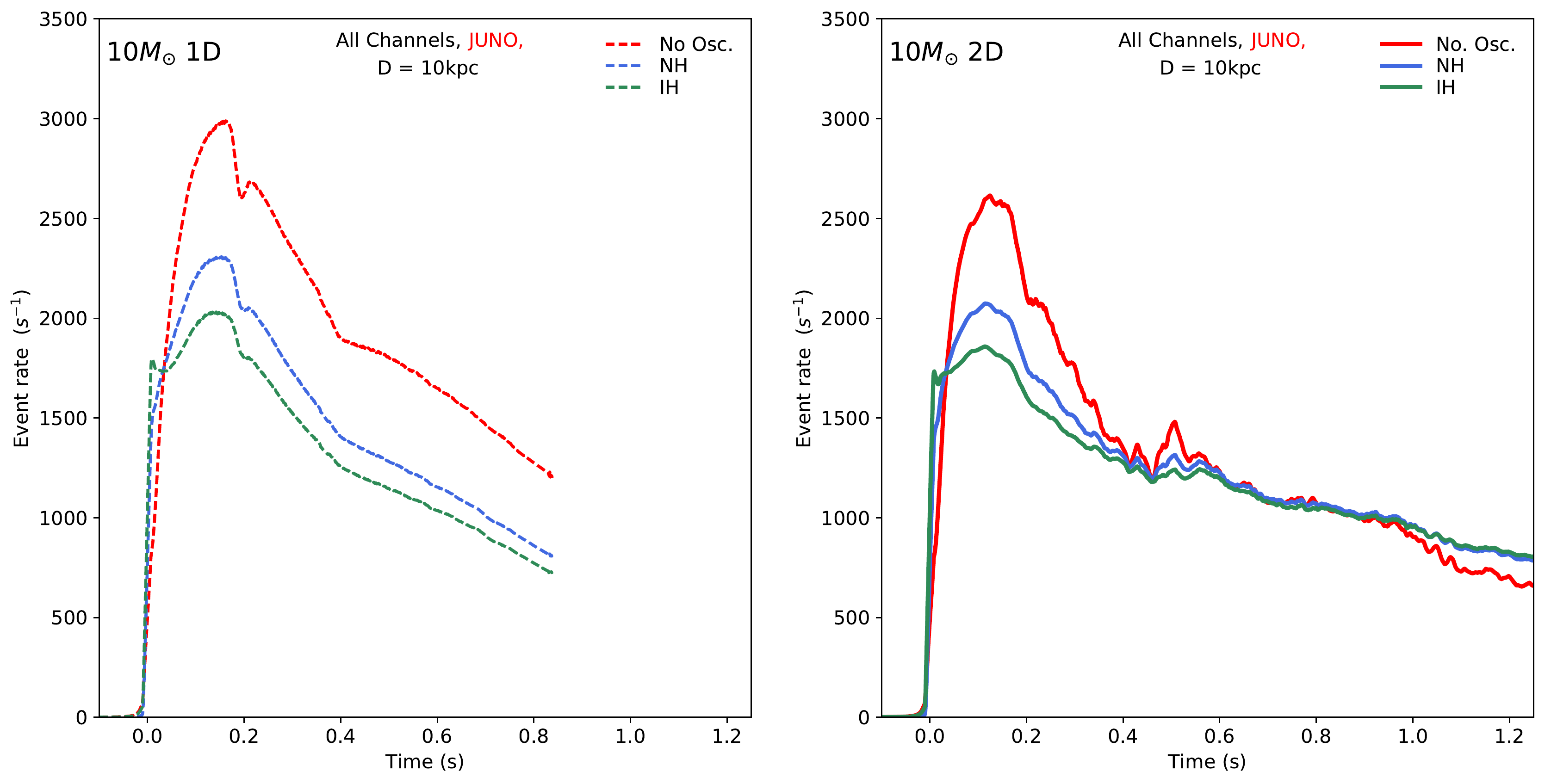}
\includegraphics[width=0.95\textwidth, angle = 0]{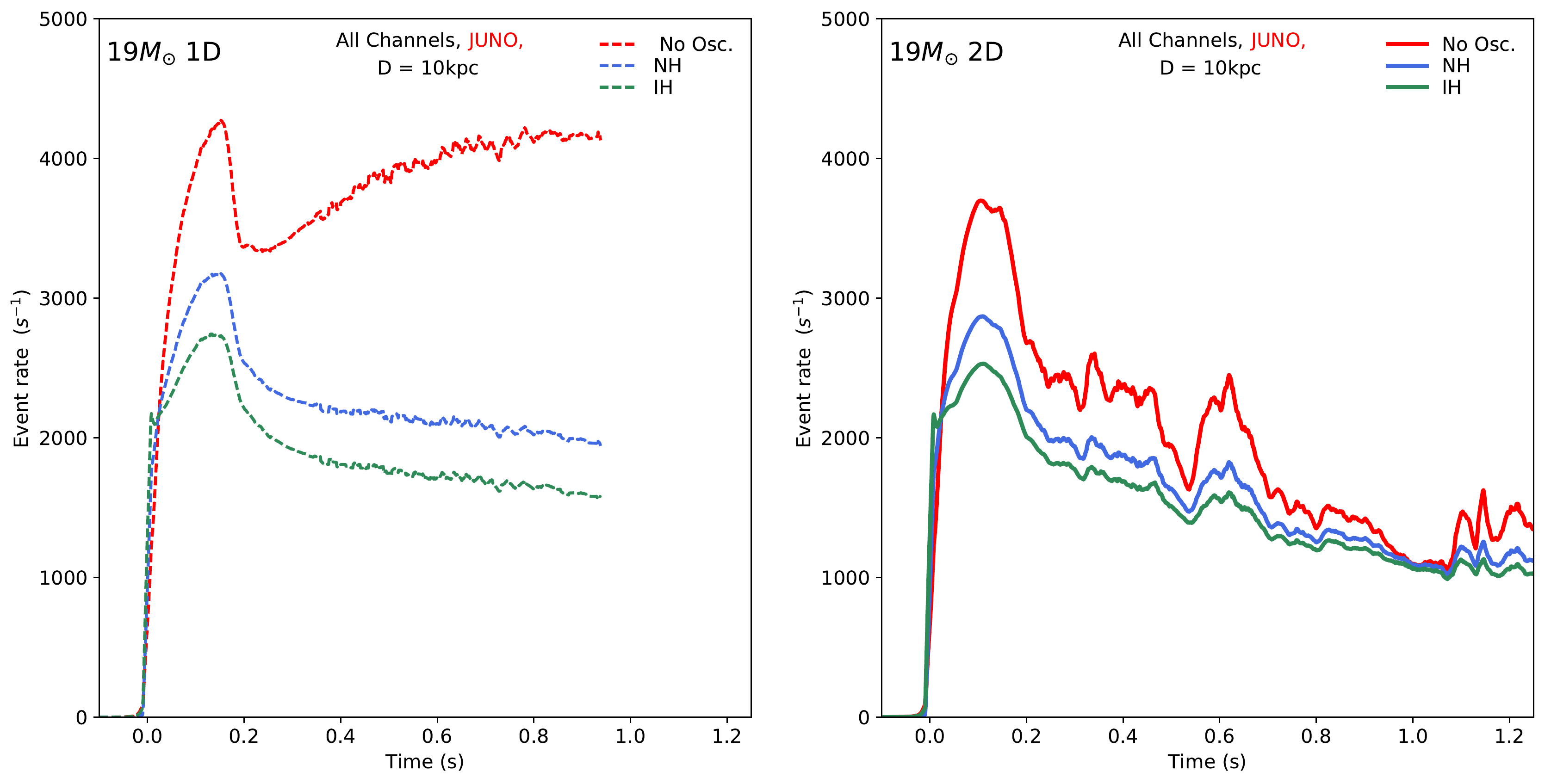}
\caption{The total, all-channel, event rates (in units of s$^{-1}$) in {\bf JUNO} at 10 kiloparsecs 
for all three oscillation models in 1D (left) and 2D (right) for the 10-M$_{\odot}$ progenitor model 
(top two plots) and the 19-M$_{\odot}$ progenitor model (bottom two plots).
As in Figure \ref{events_superk_10_19_allosc}, the 1D models are all plotted as dashed, 
while the 2D models are all solid.
\label{events_juno_10_19_allosc}}
\end{figure}

\begin{figure}
\includegraphics[width=0.95\textwidth, angle = 0]{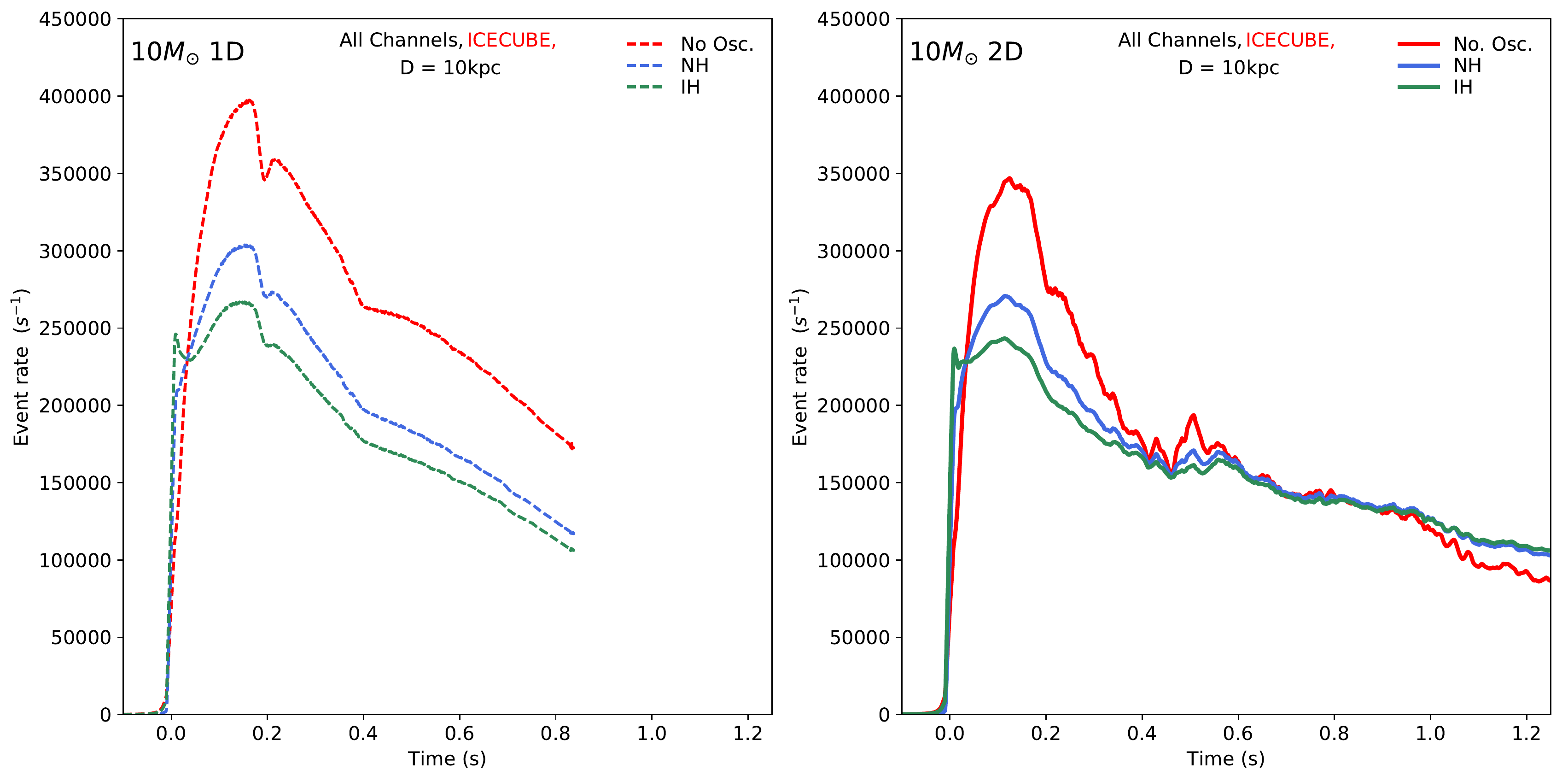}
\includegraphics[width=0.95\textwidth, angle = 0]{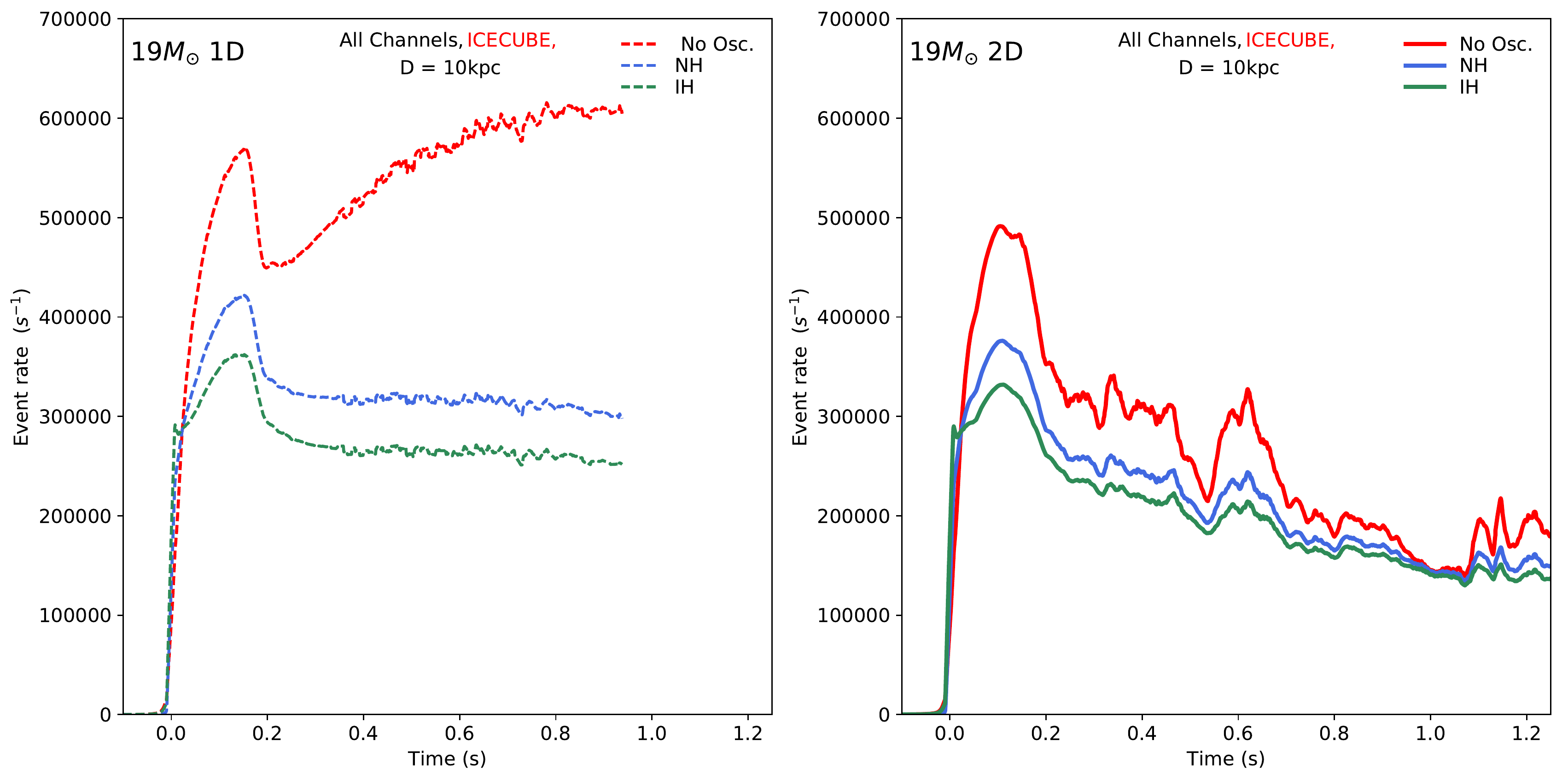}
\caption{The total, all-channel, event rates (in units of s$^{-1}$) in {\bf IceCube} at 10 kiloparsecs 
for all three oscillation models in 1D (left) and 2D (right) for the 10-M$_{\odot}$ progenitor model 
(top two plots) and the 19-M$_{\odot}$ progenitor model (bottom two plots).
As in Figure \ref{events_superk_10_19_allosc}, the 1D models are all plotted as dashed, 
while the 2D models are all solid.
\label{events_icecube_10_19_allosc}}
\end{figure}

Figures \ref{events_superk_10_19_allosc}, \ref{events_dune_10_19_allosc},
\ref{events_juno_10_19_allosc}, and \ref{events_icecube_10_19_allosc} portray
the total all-channel event rates in the Super-K, DUNE, JUNO, and IceCube detectors, respectively,
for the 10 M$_{\odot}$ and 19 M$_{\odot}$ models, for both 1D and 2D realizations, and for all neutrino
oscillation assumptions. Such plots allow one to focus on both the differences between
low-mass and higher-mass light curves and on the differences as a function of neutrino
hierarchy. 

Since Super-K is most sensitive to $\bar{\nu}_e$s, as Figure 
\ref{events_superk_10_19_allosc} suggests, the high signal rate
in the NO (no-oscillation) case is muted in either oscillation model. For the 1D (non-exploding) case,
the differences in event rate can be large, as much as $\sim$40\% for the low-mass progenitor
and more than a factor of two for the higher-mass progenitor, but for the two 2D (exploding) models
the differences between the various oscillation models in the corresponding light curves are smaller.
For the exploding (2D) models, unlike in the non-exploding models (particularly for 
the higher progenitor mass) the largest differences between the various oscillation
assumptions are at earlier post-bounce times within a few hundred milliseconds. At the 
later times, the NH and IH signal rates get closer.  As Figure \ref{events_superk_10_19_allosc}
suggests, during the first hundreds of milliseconds, there is generically an accretion 
lumninosity hump that evinces a large post-bounce and post-breakout signal rate.  
For the low-mass progenitor this hump in 1D decays quickly (with an ``e-folding" 
time near $\sim$0.8 seconds), but for the higher-mass progenitor 
the 1D signal drops and then rises or flattens.  For this 19-M$_{\odot}$ progenitor, 
the drop is due to the accretion of the silicon/oxygen
interface and the associated drop in mass flux at the shock, while the rise for the 
NO case and the near-flattening for the NH and IH cases are due to the fact 
that accretion continues in a non-exploding model and the average and RMS neutrino energies 
continue to grow (see Figures \ref{lum_rms_noosc}, \ref{lum_rms_nh}, and \ref{lum_rms_ih}) 
as the core contracts and the neutrinospheres compressionally heat.  The neutrino-matter
cross sections in detectors increase with neutrino energy (Figure \ref{cross}). So, for 
higher-mass exploding (2D) models, the accretion of the silicon/oxygen interface 
results in a decrease in the event rate, but this decrease is less manifest due 
to the almost simultaneous explosion and reversal of net accretion.
For the lower-mass progenitors, the magnitude of the density drop at this interface is smaller
and explosions can occur before it, or a similar structure, is accreted (Radice et al. 2017).
Hence, the accretion powered phase, the accretion of the silicon/oxygen interface itself,
the progenitor-model and/or compactness (seen mostly through the overall event rate), and 
the explosion/no-explosion dichotomy are all in principle observables (Horiuchi et al. 2017).  
The NH and IH hierarchies are less easily discerned in Super-K, though the earlier 
breakout burst is rather different for the two scenarios (Wallace et al. 2016).

On the other hand, DUNE (Figure \ref{events_dune_10_19_allosc}), sensitive 
as it is to $\nu_e$s, manifests at early post-bounce times an inversion 
in the event rate ordering for the NO, NH, and IH situations
vis \`a vis Super-K for both the 10-M$_{\odot}$ and 19-M$_{\odot}$.  In Super-K, the early 
event rates around the hump are in the order NO $>$ NH $>$ IH for both non-exploding and 
exploding models, but in DUNE they are NH $>$ IH $>$ NO.  After $\sim$200 milliseconds,
in both the 10-M$_{\odot}$ and 19-M$_{\odot}$ non-exploding models this order reverses, 
and is NO $>$ IH $>$ NH.  So, for the non-exploding models and at later times, the NH/IH 
order reverses in DUNE relative to the behavior in Super-K.  For the exploding 
(2D) models, the event rate differences between the hierarchies are smaller.  However, we note
that at the earlier times ($<$200 milliseconds after bounce) the light curve shapes
in DUNE and Super-K are very different, reflecting their differential neutrino flavor 
sensitivities.  This, along with a potential capacity in each detector
to discriminate channel, flavor, and spectra, could be used to derive both oscillation
modality and progenitor-model constraints. We emphasize, as do Wallace et al. (2016), 
and as is suggested via Figure \ref{events_dune_10_19_allosc} and Figures \ref{events_dune_all} and \ref{fig14} 
below, that DUNE's $\nu_e$ sensitivity makes it the best detector of the four highlighted
in this paper to discern the breakout burst, and with it to distinguish the normal 
from the inverted hierarchy.  

As Figures \ref{events_superk_10_19_allosc}, \ref{events_juno_10_19_allosc}, \& \ref{events_icecube_10_19_allosc} demonstrate, 
due to the $\bar{\nu}_e$ sensitivity in all three, the light curves in JUNO, IceCube, and Super-K 
are qualitatively similar. IceCube, however, if the fiducial volume   
we have assumed for it obtains, would experience a much higher rate of 
signal accumulation\footnote{At 10 kiloparsecs, 
a CCSN would register during the first second of post-bounce evolution
$\sim$10$^4$ times as many events in IceCube as were culled by Kamioka II from SN 1987A.  
The other detectors highlighted here would witness $\sim$10$^2$ times more events.}. 
Nevertheless, the temporal structure of 
the neutrino burst during the dynamical supernova phase is best discerned via IceCube,
but the flavor and spectral character of the burst can be determined clearly only via the channel
and event energy discrimination that might be possible in the collective combination of 
Super-K, DUNE, and JUNO.  The differential senstivity of Super-K and JUNO on the one hand
and DUNE on the other vis \`a vis $\bar{\nu}_e$s and $\nu_e$s will be particularly 
revealing.  As noted, we have focused in this paper on all-channel energy-integrated rates
and defer a discussion of the spectral and channel capabilities in these 
three detectors (and in Hyper-K) to a later paper. 

In Figures \ref{events_superk_all}, \ref{events_dune_all}, \ref{events_juno_all}, and
\ref{events_icecube_all}, we present all the all-channel light curves for all the progenitors
studied, for both 1D (non-exploding) and 2D (exploding) realizations in Super-K, DUNE, JUNO, and IceCube.
The basic systematics with progenitor articulated and described above using only
the 10-M$_{\odot}$ to 19-M$_{\odot}$ models survives. The wide range of signal rates 
as a function of progenitor mass and compactness is clear.  For the non-exploding models,
the early-time event rates can vary by a factor of $\sim$2, while the late-time event rates 
can vary by a factor of $\sim$10 for the non-oscillating case and $\sim$5 for the 
oscillating cases.  For the exploding models, variations of at least a factor of $\sim$2
are expected from 9-M$_{\odot}$ to 21-M$_{\odot}$. Since a galactic supernova will almost
certainly be measured electromagnetically, the distance to the supernova will be known.
Given the distance, the absolute flux and fluence can be determined and with it 
some measure of the core structure/compactness of the progenitor.  The electromagnetic data will
likely reveal the progenitor mass, so the absolute neutrino signal can be paired with
the progenitor star to constrain stellar evolution and supernova models together.

\subsection{Detection Ranges and Feature Measurement}
\label{ranges}

We emphasize that a detected signal will be in our galaxy, or in very local environs (such as
the LMC or SMC).  Given this, the distance to the supernova (and most probably the progenitor
mass and photon luminosity) will be known via its electromagnetic counterparts.
Hence, there is no distance degeneracy for exploding models.  This allows us readily
to distinguish the signal rate levels for the various progenitors and to experimentally
test the models. With a distance (and perhaps an electromagnetically-derived progenitor mass), the detection of the
neutrino signals in the various detectors described will provide excellent
constraints on and insights into supernova theory and on the models we use here (Radice et
al. 2017; Vartanyan et al. 2018a; Burrows et al. 2018).

\begin{figure}
\includegraphics[width=0.80\textwidth, angle = 0]{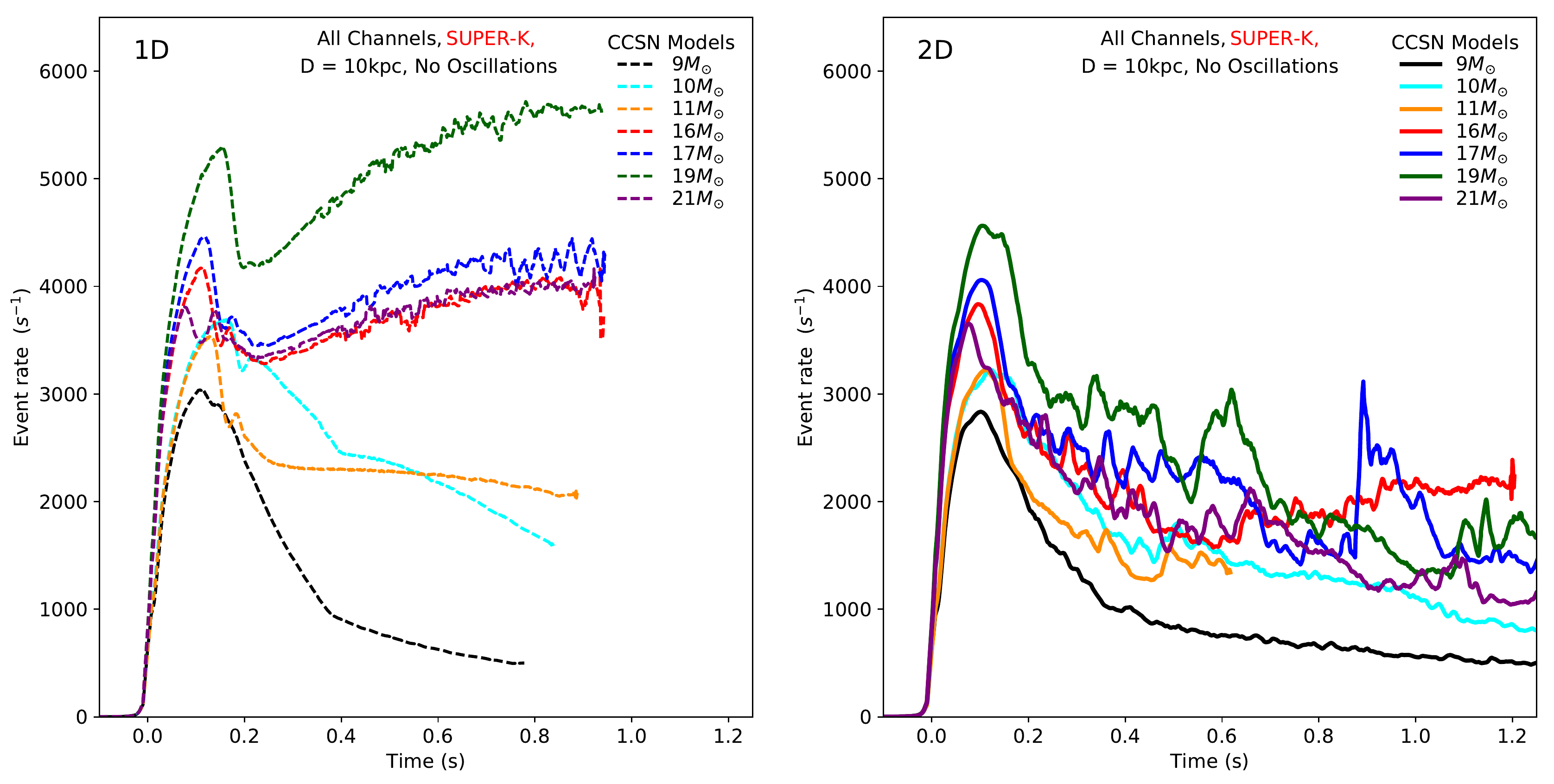}
\includegraphics[width=0.80\textwidth, angle = 0]{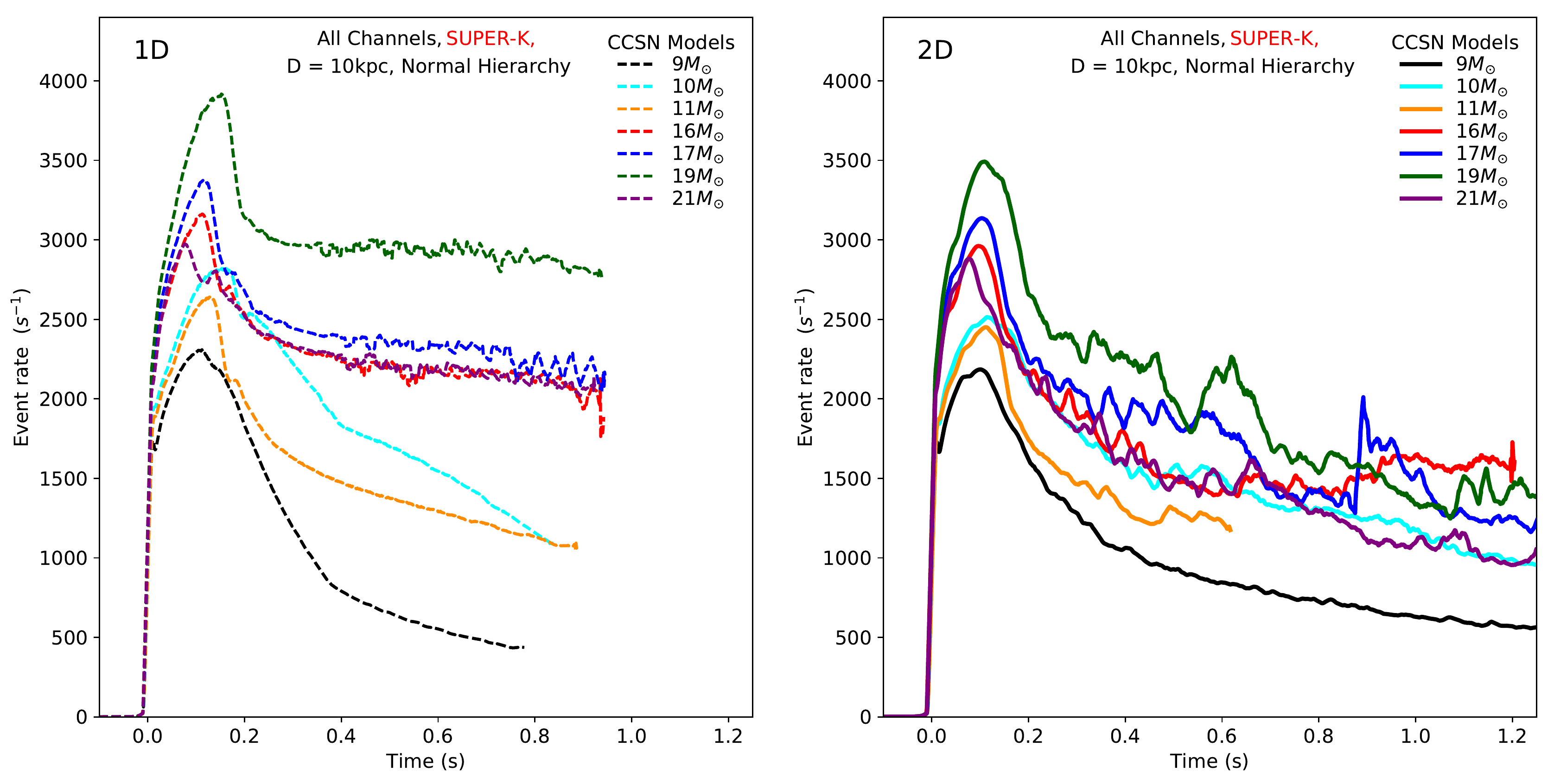}
\includegraphics[width=0.80\textwidth, angle = 0]{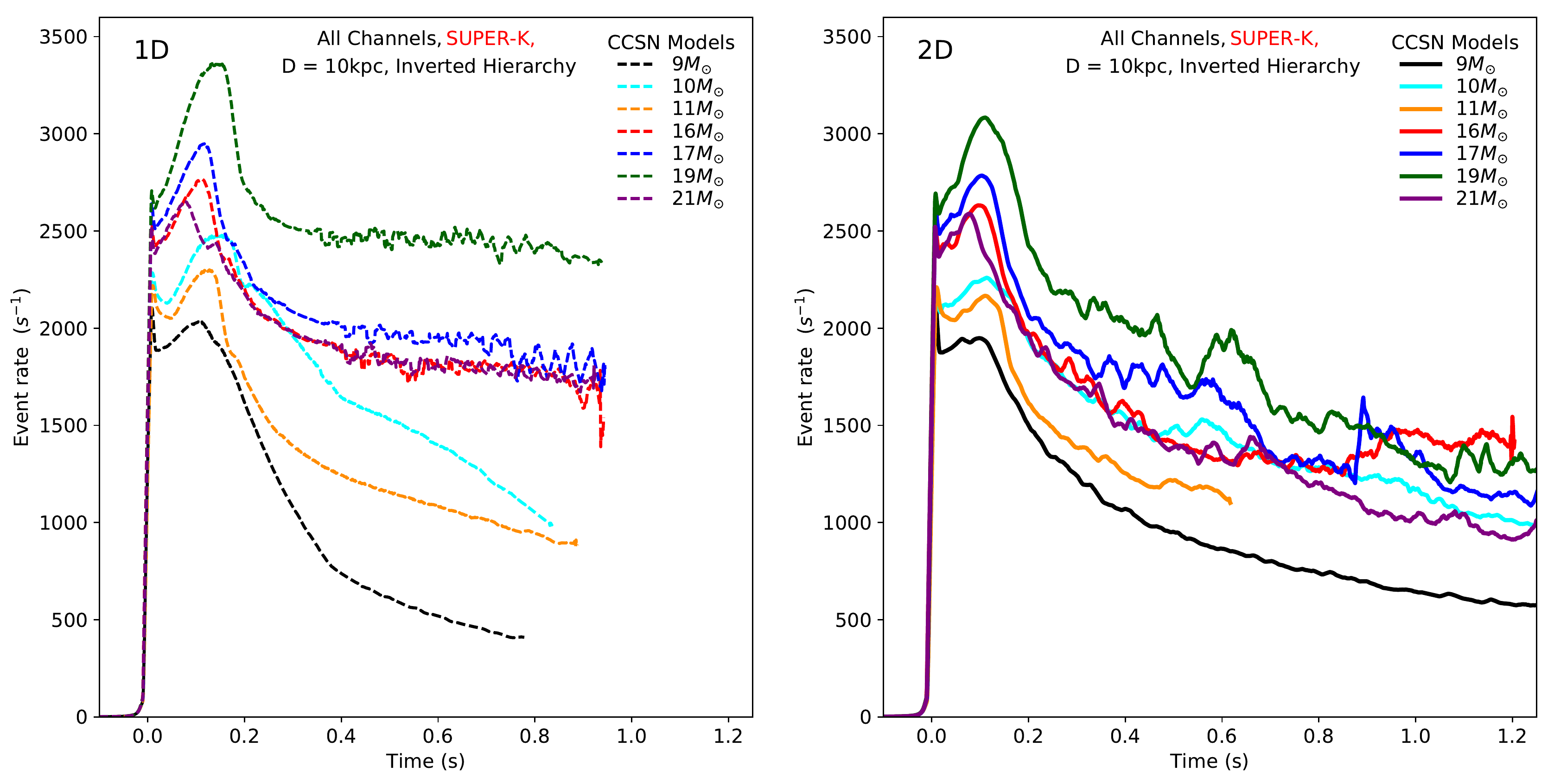}
\caption{The total, all-channel, event rates (in units of s$^{-1}$) in {\bf Super-K} at 10 kiloparsecs
in the 1D (non-exploding) and 2D (exploding) calculations for all the progenitor models 
(9, 10, 11, 16, 17, 19, and 21 M$_{\odot}$) addressed in this study.  The left plots are 
for the 1D models (in this paper, non-exploding) and the right plots are for the 2D models (exploding).
The non-oscillation cases are in the top pair of figures, the normal hierarchy results are depicted in
the middle pair, and the inverted hierarchy rresults are shown in the bottom pair (again, 1D [left], 2D [right]).
As above in Figures \ref{events_superk_10_19_allosc} through \ref{events_icecube_10_19_allosc},
the 1D models are all plotted as dashed, while the 2D models are all solid.
\label{events_superk_all}}
\end{figure}

\begin{figure}
\includegraphics[width=0.80\textwidth, angle = 0]{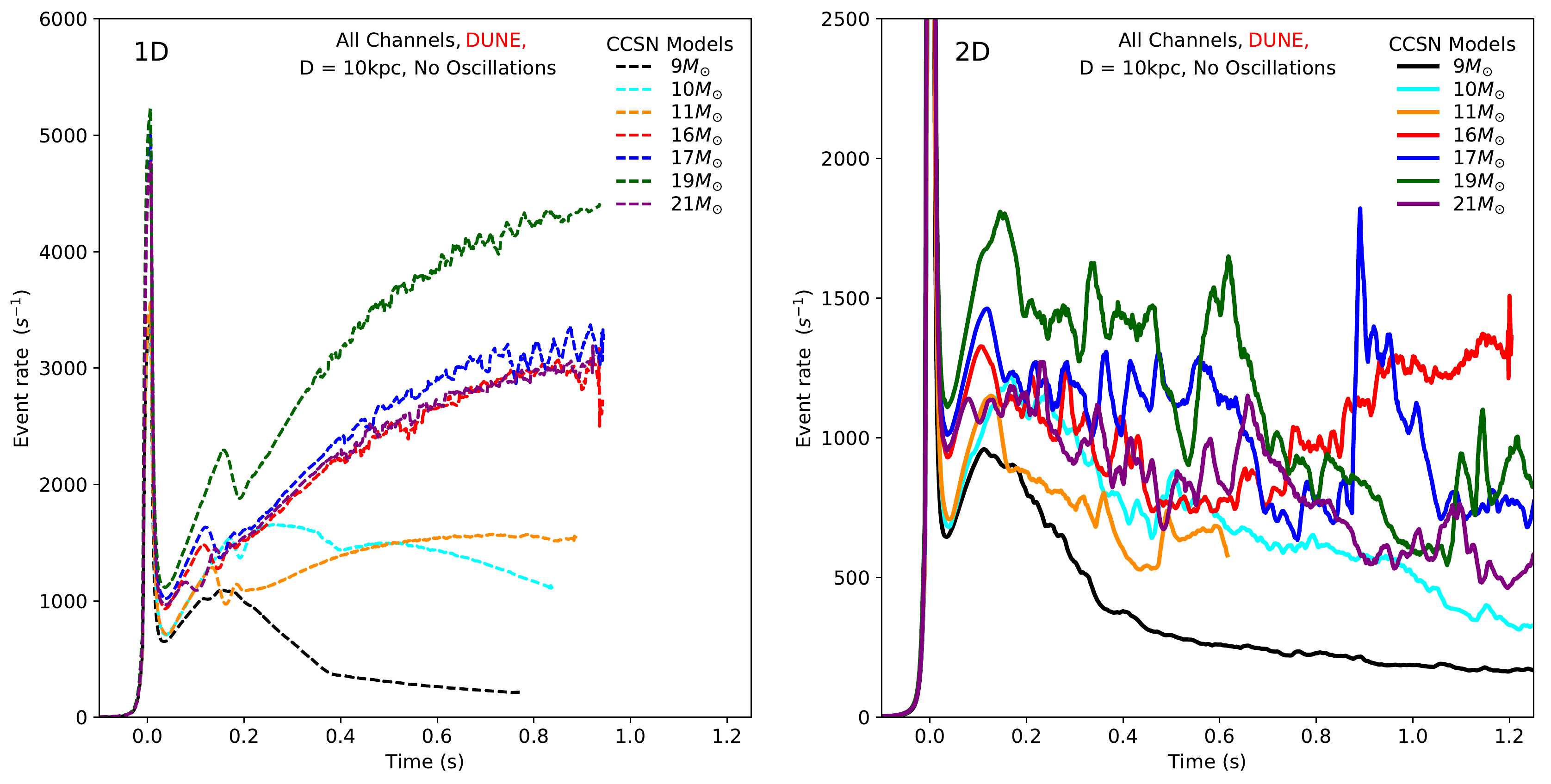}
\includegraphics[width=0.80\textwidth, angle = 0]{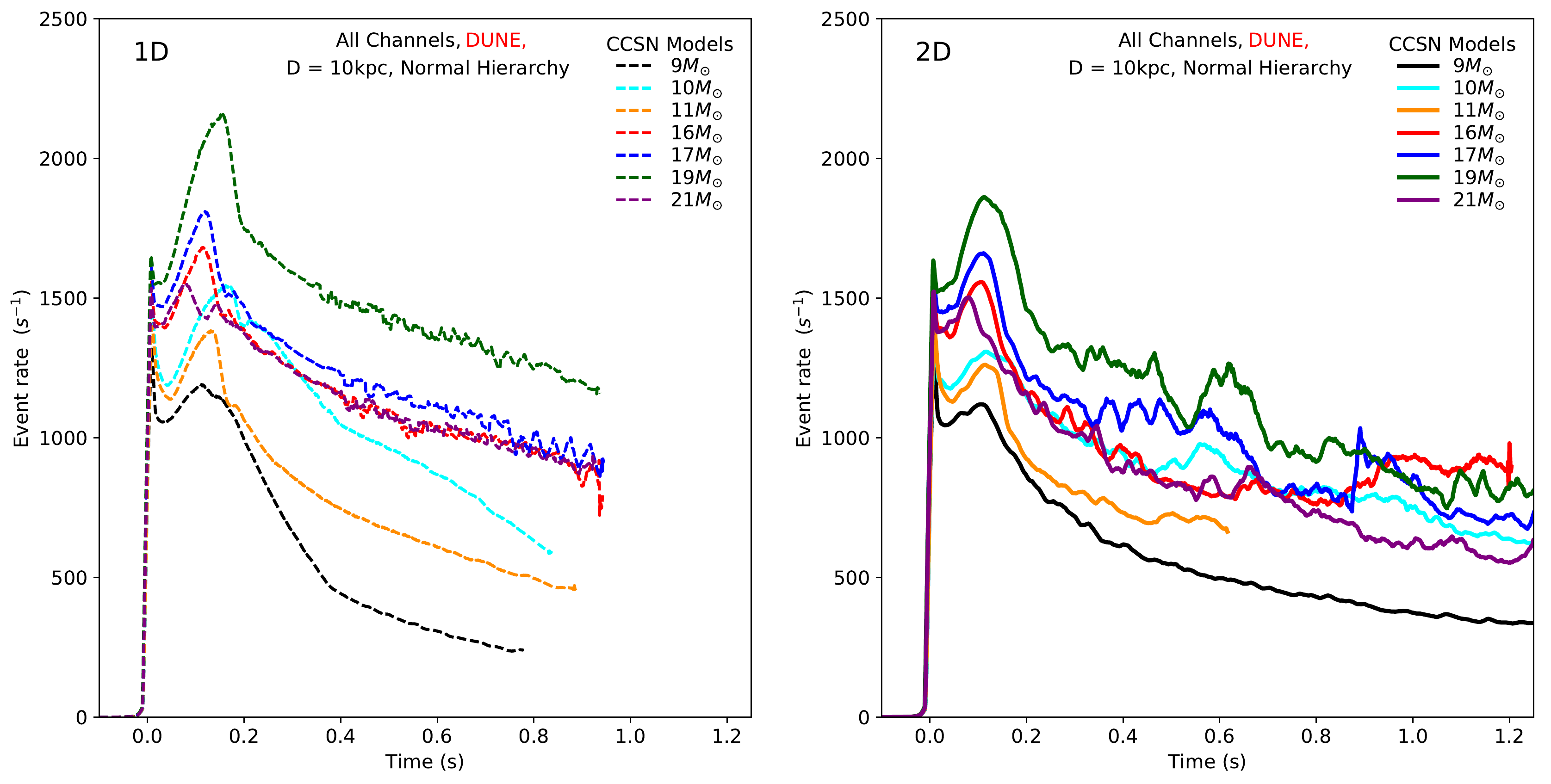}
\includegraphics[width=0.80\textwidth, angle = 0]{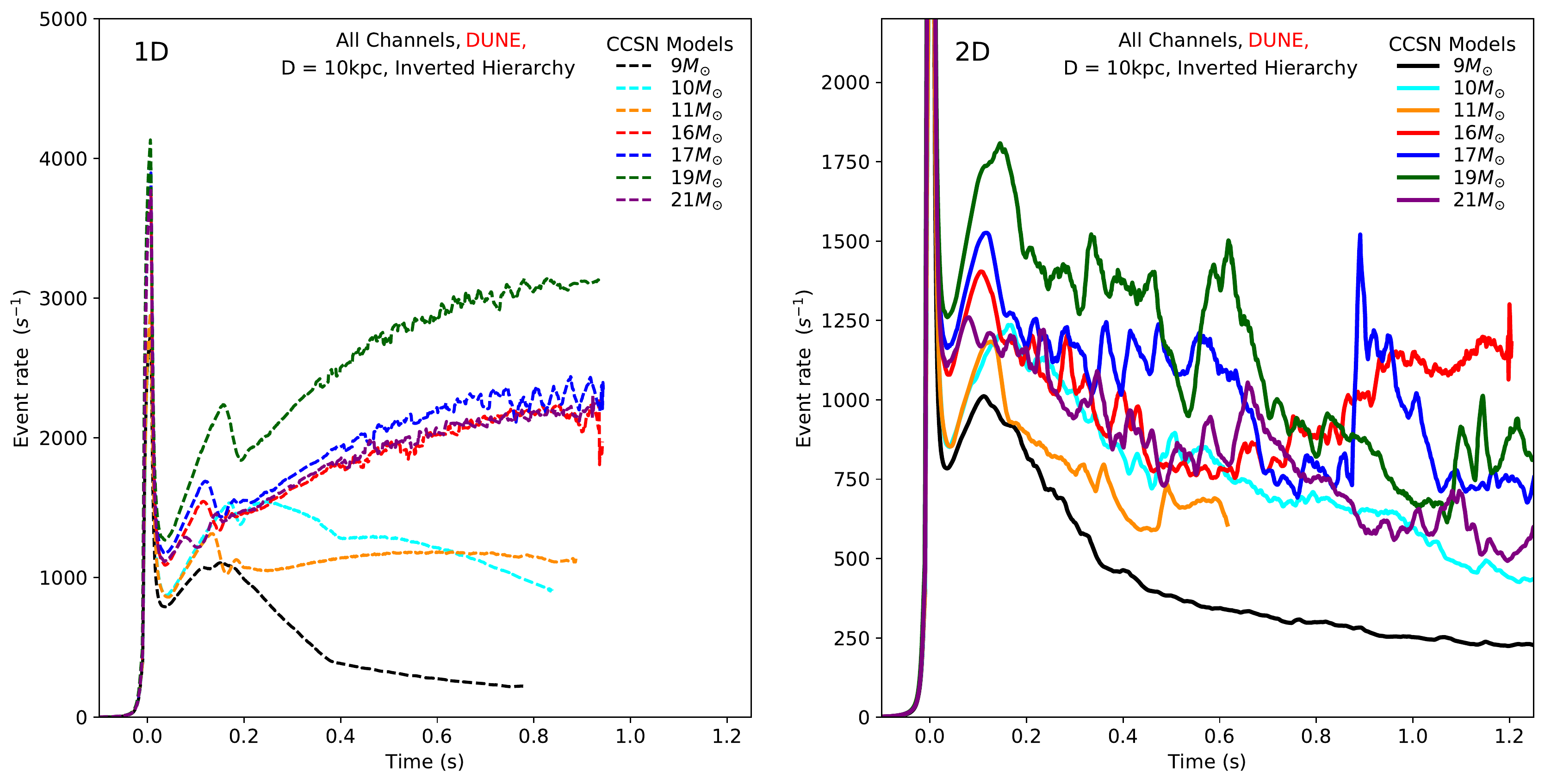}
\caption{The same as Figure \ref{events_superk_all}, but for {\bf DUNE}.
\label{events_dune_all}}
\end{figure}

\begin{figure}
\includegraphics[width=0.80\textwidth, angle = 0]{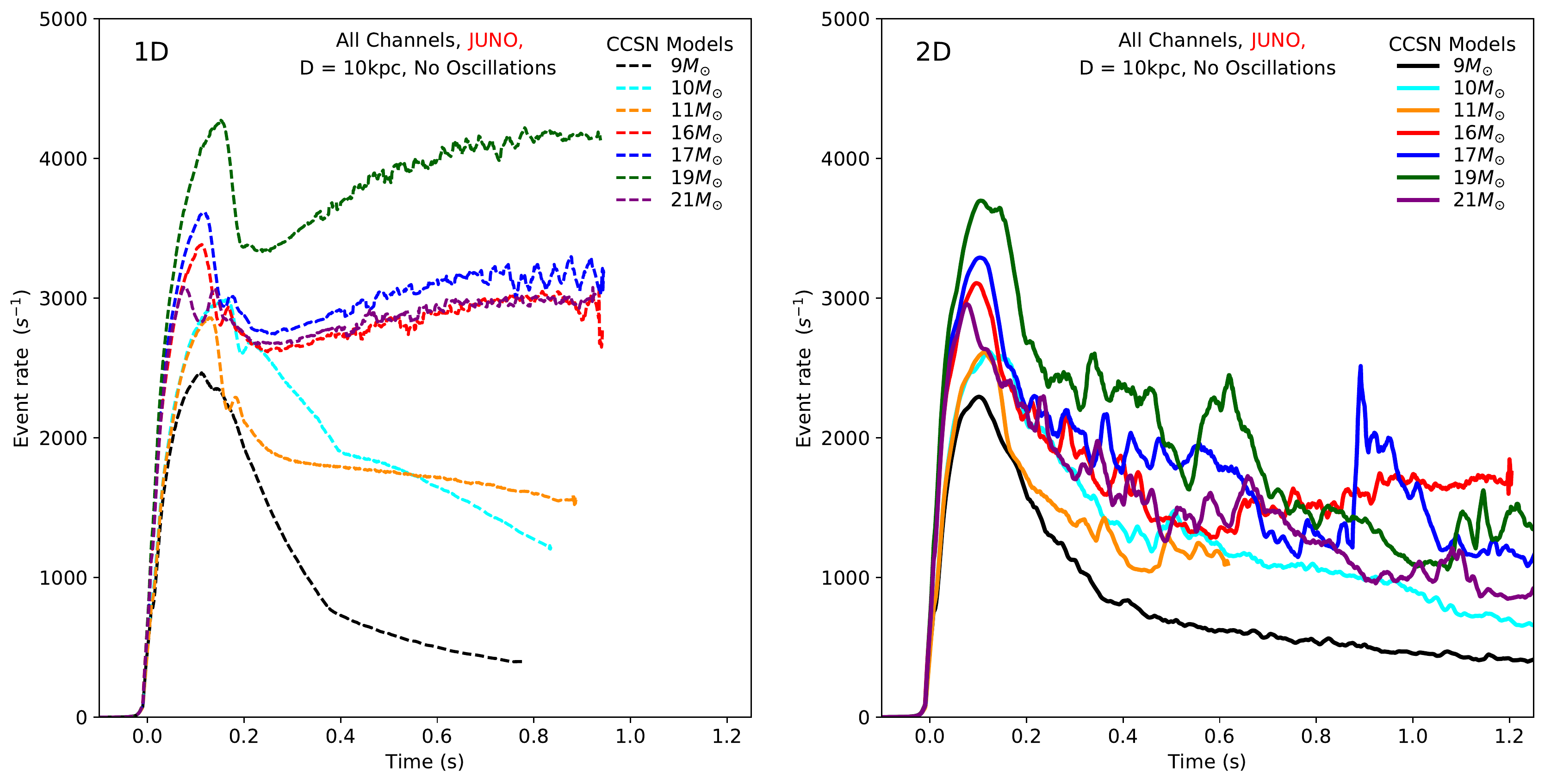}
\includegraphics[width=0.80\textwidth, angle = 0]{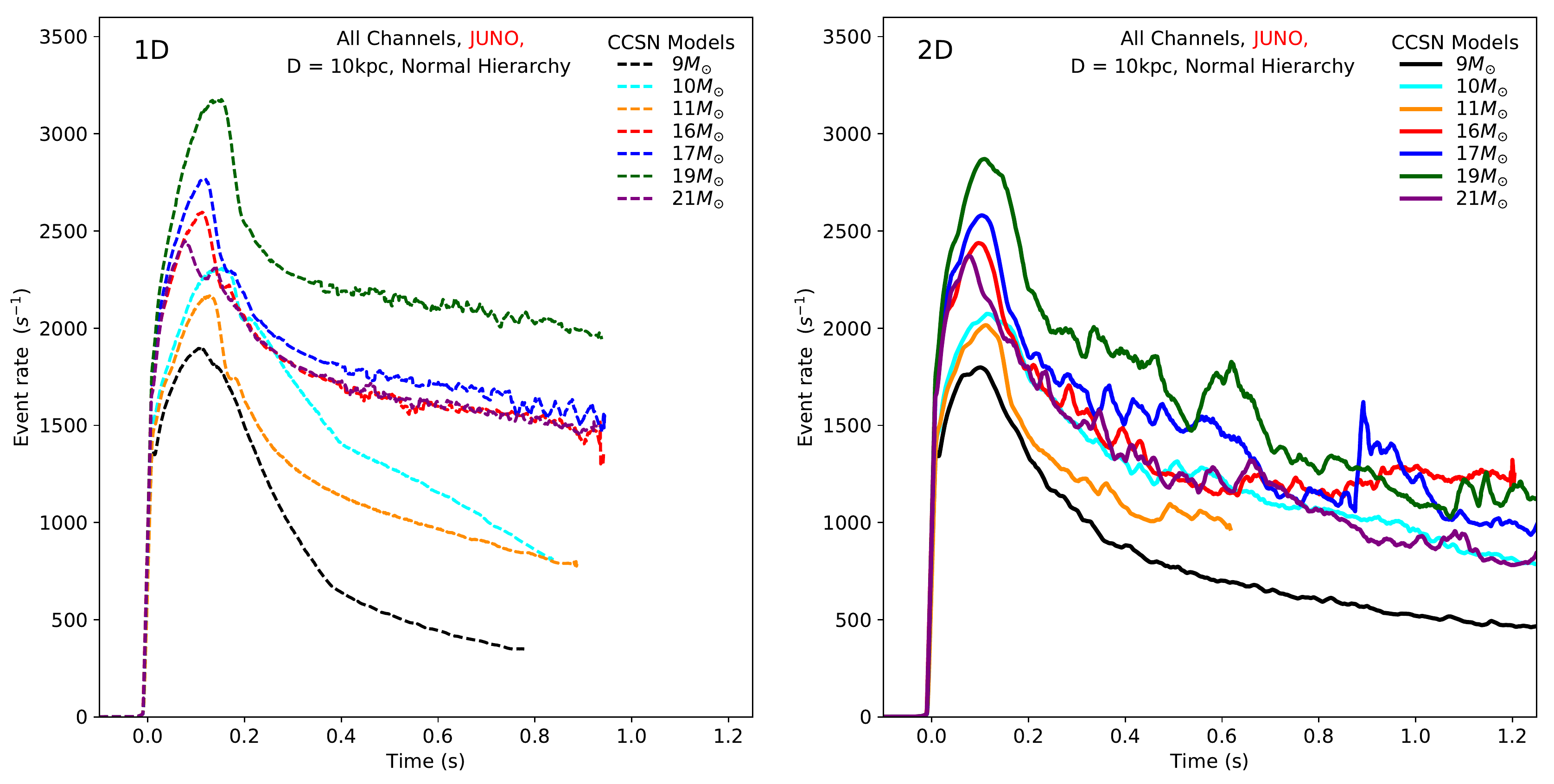}
\includegraphics[width=0.80\textwidth, angle = 0]{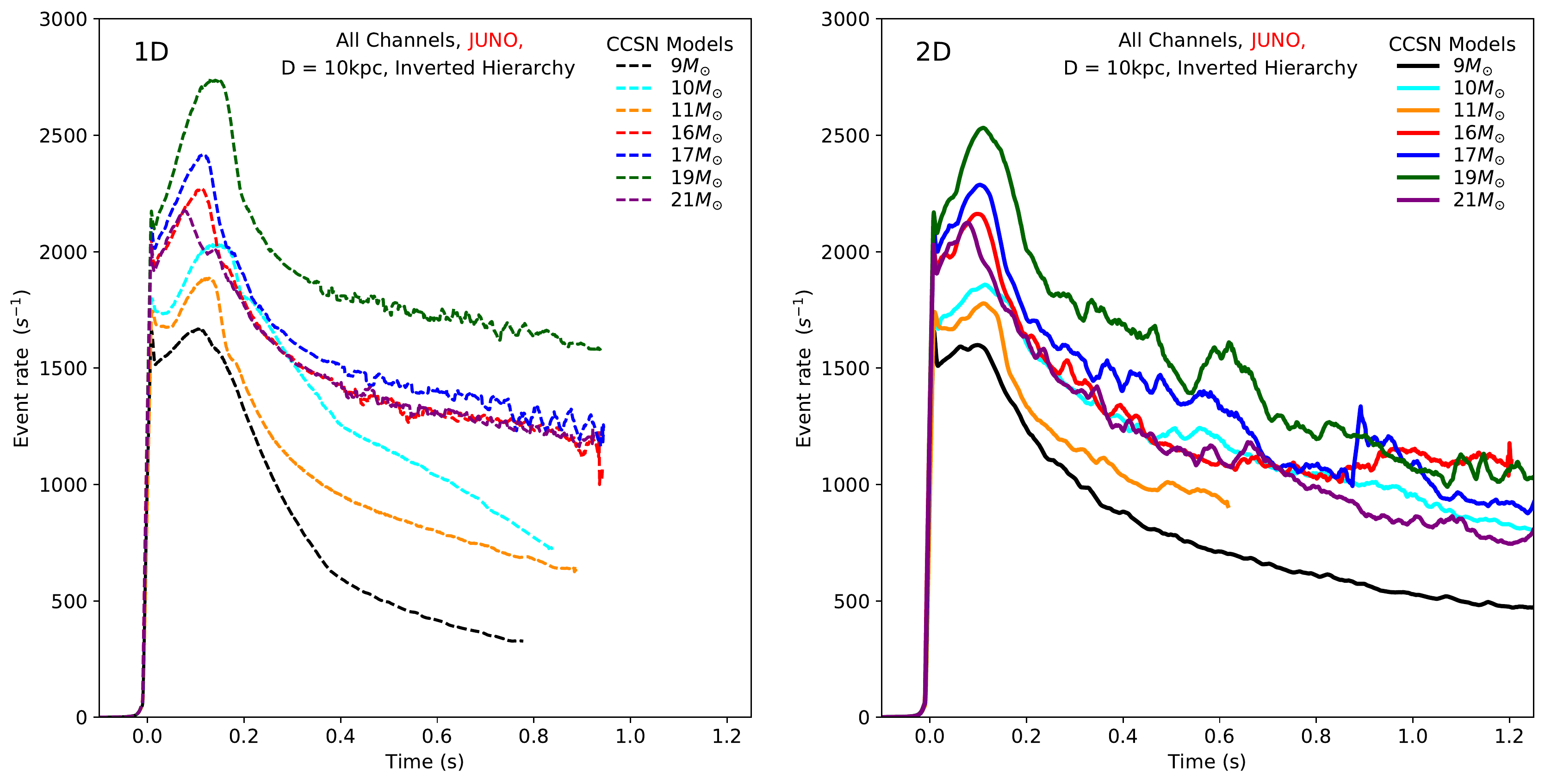}
\caption{The same as Figure \ref{events_superk_all}, but for {\bf JUNO}.
\label{events_juno_all}}
\end{figure}

\begin{figure}
\includegraphics[width=0.80\textwidth, angle = 0]{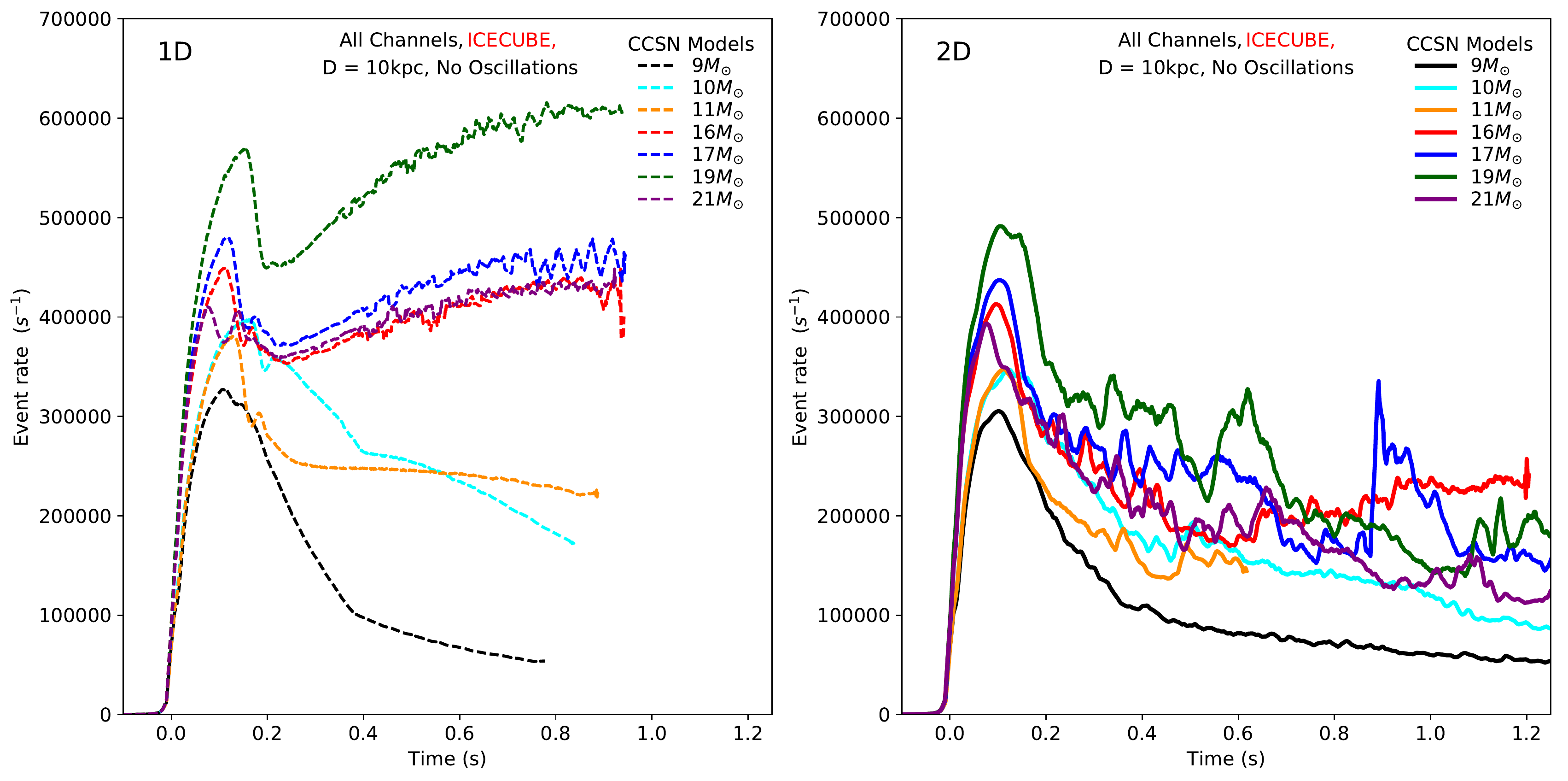}
\includegraphics[width=0.80\textwidth, angle = 0]{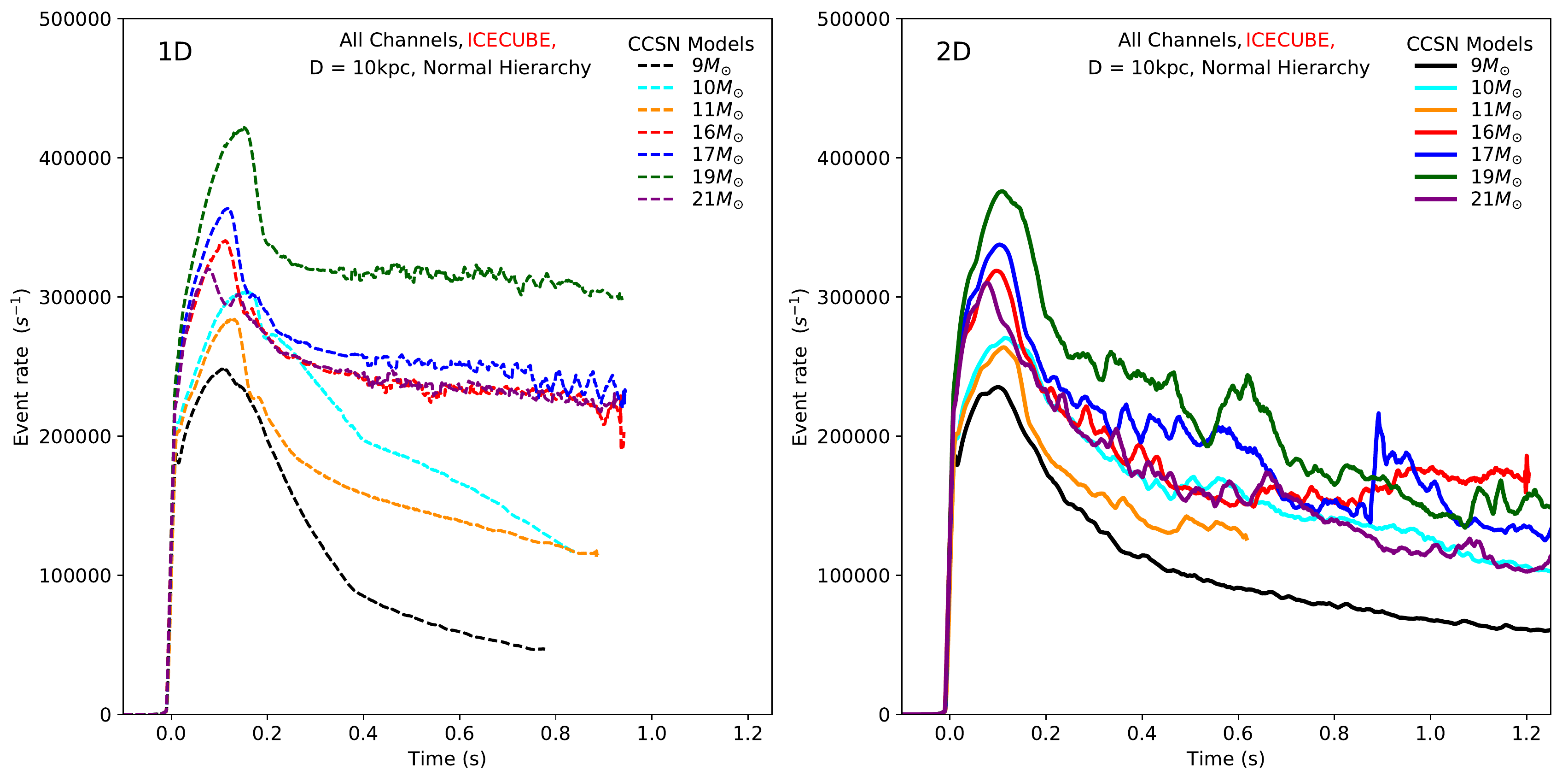}
\includegraphics[width=0.80\textwidth, angle = 0]{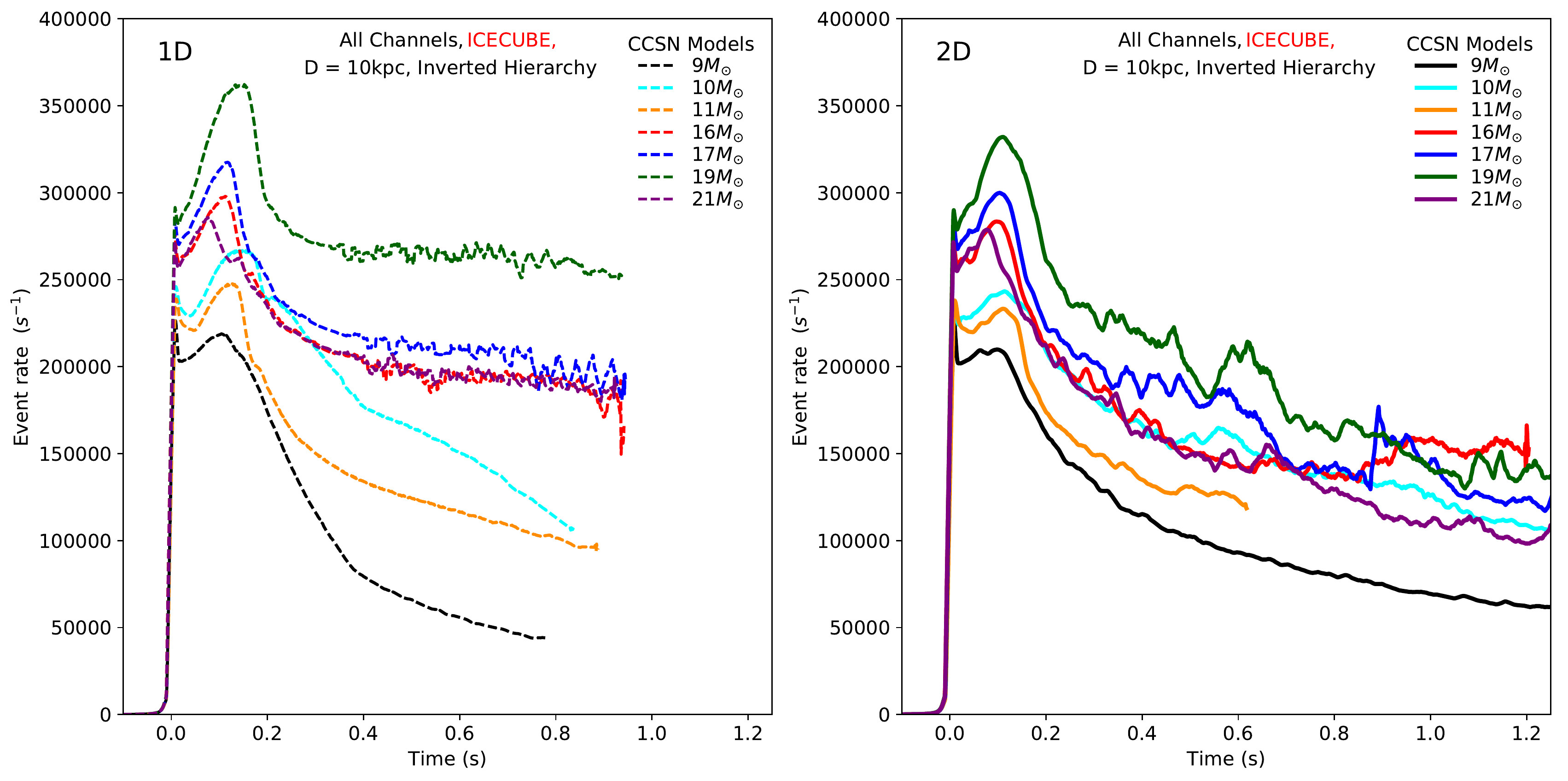}
\caption{The same as Figure \ref{events_superk_all}, but for {\bf IceCube}.
\label{events_icecube_all}}
\end{figure}

Table \ref{tab:PPer} provides the total number of events expected at 10 kiloparsecs
in all the models and for all the detectors studied.  For Super-K, DUNE, and JUNO the effective background
count rate is negligible during times of order one second.  Therefore, if we assume Poisson statistics
for the signal fluctuations, deriving the total number of events to $\sim$10\% accuracy at 1-$\sigma$ would require
that $\frac{1}{\sqrt{N}}$ $\sim$ 0.1, where $N$ is the number of events.  This criterion yields $N \sim 100$,
and, therefore, a range of $\sim$100 kiloparsecs for these three detectors. This encompasses
the entire Milky Way and the Magellanic Clouds.  However, an accuracy of 10\% for the entire signal
during the dynamical ``first second" is not that impressive.  Astronomers would want to discern 
various temporal structures, such as the post-bounce accretion hump. Using Figures 
\ref{events_superk_10_19_allosc}$-$\ref{events_juno_10_19_allosc} and \ref{events_superk_all}$-$\ref{events_juno_all},
we can determine that to capture the number fluence during the accretion hump (assuming that it lasts for 
$\sim$200 milliseconds) to at least $\sim$10\% accuracy the distance of the supernova needs to be closer 
than 20 to 25 kpc for Super-K and $\sim$10 to 15 kpc for DUNE, depending upon oscillation and progenitor model.
This still encompasses most of the Milky Way.  The corresponding numbers for JUNO are in most cases in between.
In general, and approximately, in order to discern a temporal feature, ``$i$", of width 
$\Delta{t}$ whose average signal rate is $\dot{E_i}$ ($=\frac{dN_{det}}{dt}$) to within a 
fraction, $f$, requires that $f > N_i^{-{1/2}} \sim \frac{1}{(\dot{E_i}\Delta{t})^{1/2}}$.
For example, to identify a temporal fluctuation in the event rate of 10\% that might occur over a $\Delta{t}$ of 10 milliseconds
requires a ``background" $\dot{E}_i$ of $\sim$10,000 Hertz. In Super-K, the higher-mass 
progenitors need to be within $\sim$5$-$6 kpc to achieve this, depending upon oscillation model and epoch
of fluctuation.  Such fluctuations might arise due to episodic accretion or rotation and would usefully 
constrain the character of the turbulence in the core and behind the shock wave. A feature with a larger $\Delta{t}$
would be discernible out to a correspondingly greater distance.  In fact, the range out to which one
could discern a fluctuation $f$ of duration $\Delta{t}$ is proportional to $f$ and $\sqrt{\Delta{t}}$. 
Hence, to capture a small fluctuation of 1\% over an interval of 10 milliseconds would require a supernova 
distance below $\sim$1 kpc for most models.  Only of order 1\% of the galaxy is within this distance.

Nevertheless, as Figures \ref{events_superk_10_19_allosc}$-$\ref{events_juno_10_19_allosc} and \ref{events_superk_all}$-$\ref{events_juno_all}
clearly demonstrate, the general evolution of the neutrino light curve during the crucial dynamical supernova 
phase will be accessible for a core-collapse supernova going off in most of our galaxy. 
Moreover, as Figures \ref{events_superk_all}$-$\ref{events_juno_all} reveal, the range of event rates as a function of 
progenitor can vary by a factor of two to five, depending upon whether the model explodes and, to some degree,
upon whether the in Super-K, DUNE, and JUNO for supernovae throughout the galaxy.  
For instance, for the normal hierarchy and non-exploding models, the 9-M$_{\odot}$ and 
10-M$_{\odot}$ progenitor predictions can be distinguished in Super-K out to $\sim$100 kpc. For the exploding
models, the variation in signal rate is approximately a factor of two and the $\sim$25\% variation from model
to model seen in this model suite can be resolved out to 50$-$100 kpc. Of course, more precise discrimination requires
closer supernovae, but the numbers for the parameter examples given are already rather impressive.  

IceCube has a significant background count rate.  The supernova would be identified by a rise 
in this background. If we assume a background of 1200 $\Delta{t}$ Hertz (\S\ref{subsec:ice}), 
then the signal to background ratio for the first second after collapse ranges from $\sim$100 to $\sim$300 at 10 kpc.
This implies that a supernova anywhere in the Milky Way could be detected in IceCube.  
Since the Poissonian noise fluctuation of the signal itself binned in 10-millisecond bins 
exceeds that in the detector for distances less than $\sim$3$-$5 kpc, the signal 
in IceCube will be detector-noise-limited for much of the galaxy.
At a signal rate ($\dot{E}$) of $\sim$10$^5$ Hertz, the signal-to-background ratio
is $\sim$9 for $\Delta{t} = 10$ milliseconds and $\sim$4 for $\Delta{t} = 2$ milliseconds. At 
an $\dot{E}$ of $\sim$$3\times 10^5$, the corresponding numbers are $\sim$27 and $\sim$12.
Therefore, at the higher rate (see Figure \ref{events_icecube_all}), IceCube should be able 
to measure the event rate in 10-millisecond bins to $\sim$4\%, and in 20-millisecond bins 
to $\sim$3\%.  This would enable exquisite model discrimination at 10 kpc. 
Within a given oscillation paradigm, IceCube is able to discriminate the various progenitors for 
any galactic CCSN solely on the basis of the general level of flux measured.  Hence,
with $\sim$10$^2$ times the signal rate found in the other detectors, it should be able to achieve
for a time bin of 10 milliseconds and distances less than $\sim$5 kpc (approximately interior to which the 
signal noise exceeds the detector noise) $\sim$10 times the precision in the estimation of instantaneous 
event rate (ten times smaller $f$s). Hence, IceCube will be able to provide exquisite 
all-channel neutrino light curves throughout most of the galaxy.

\begin{figure}
\includegraphics[width=0.95\textwidth, angle = 0]{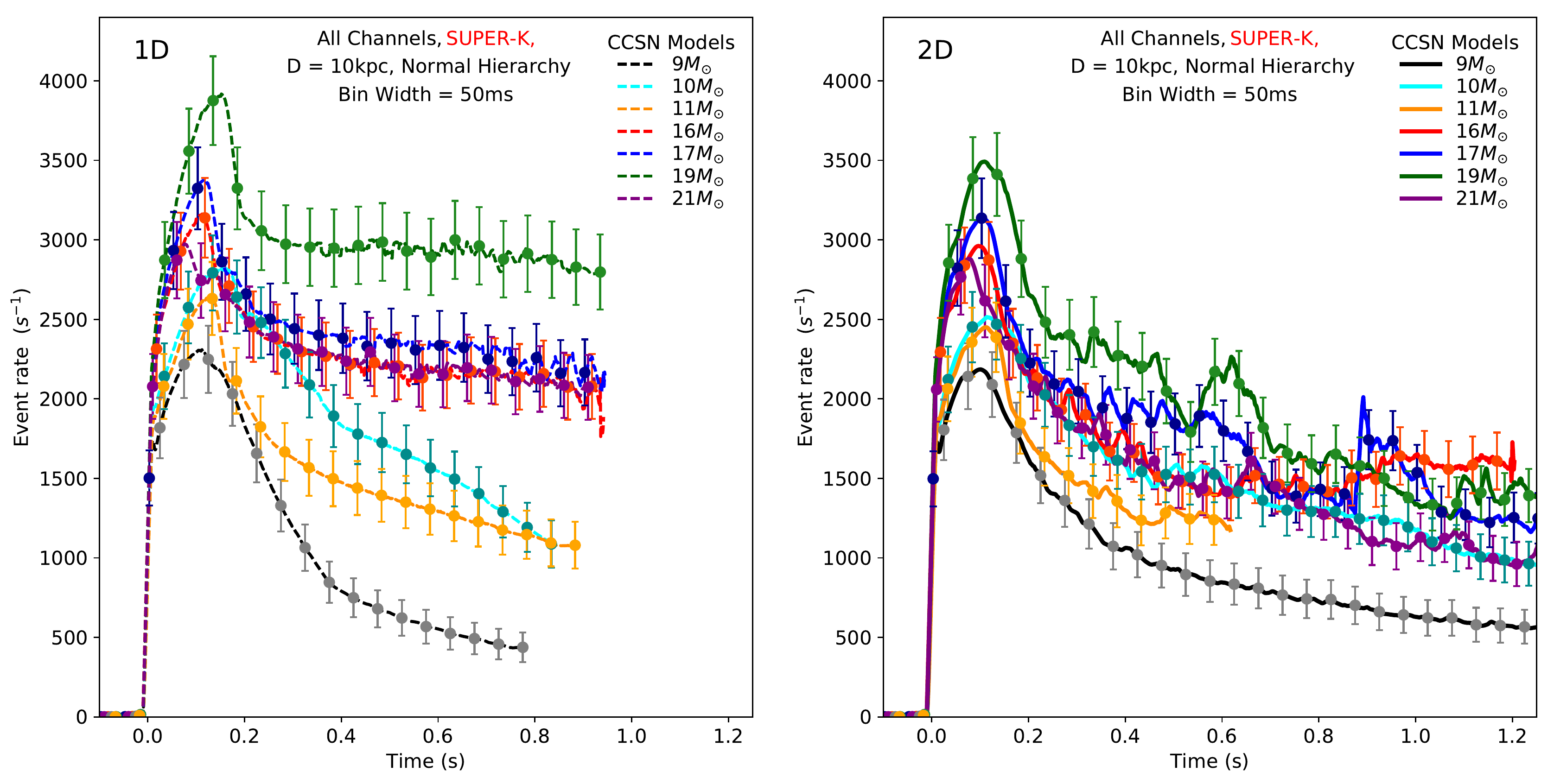}
\includegraphics[width=0.95\textwidth, angle = 0]{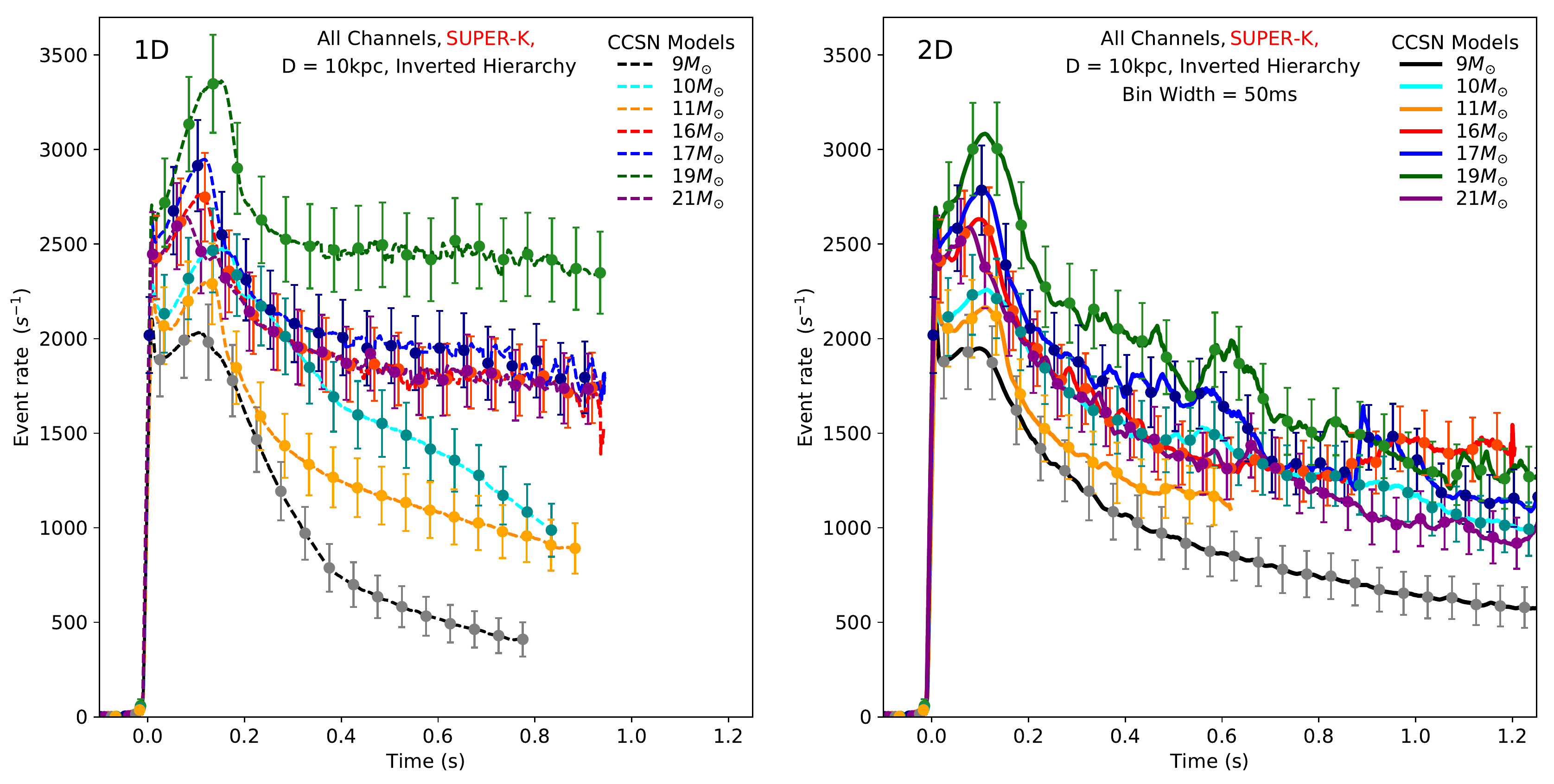}
\caption{Similar to what is plotted in Fig. \ref{events_superk_all}, 
this plot reprises the total event rates (in units of s$^{-1}$) in 
{\bf Super-K} at 10 kiloparsecs and versus time after bounce, 
assuming the normal hierarchy (top) and inverted hierarchy (bottom) 
for all the 1D and 2D progenitor models highlighted in this paper.  
Also included, however, are error bars in these signal rates assuming 
50-millisecond time bins and Poissonian noise. See text for a discussion.
\label{fig13}}
\end{figure}

\begin{figure}
\includegraphics[width=0.95\textwidth, angle = 0]{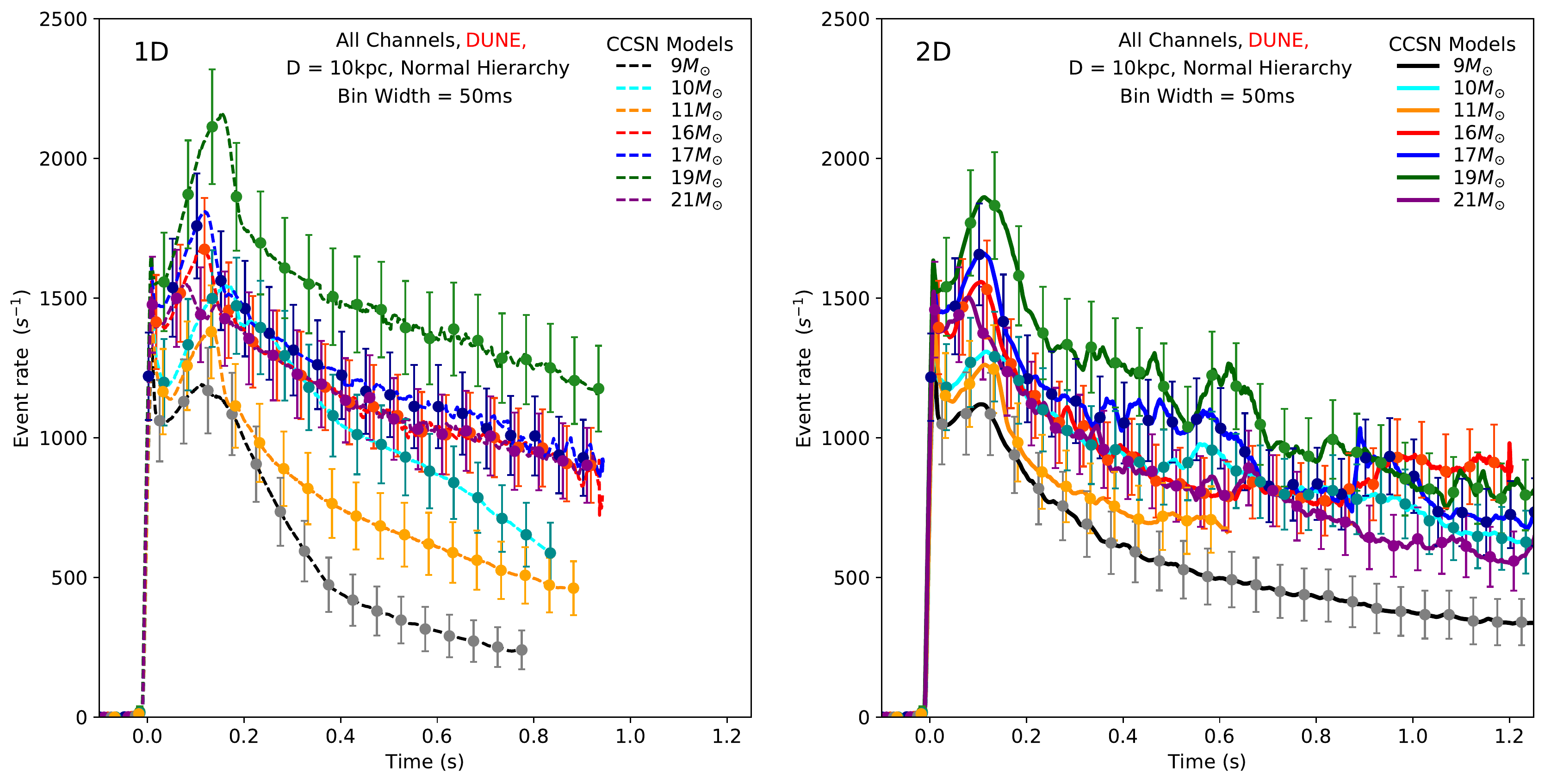}
\includegraphics[width=0.95\textwidth, angle = 0]{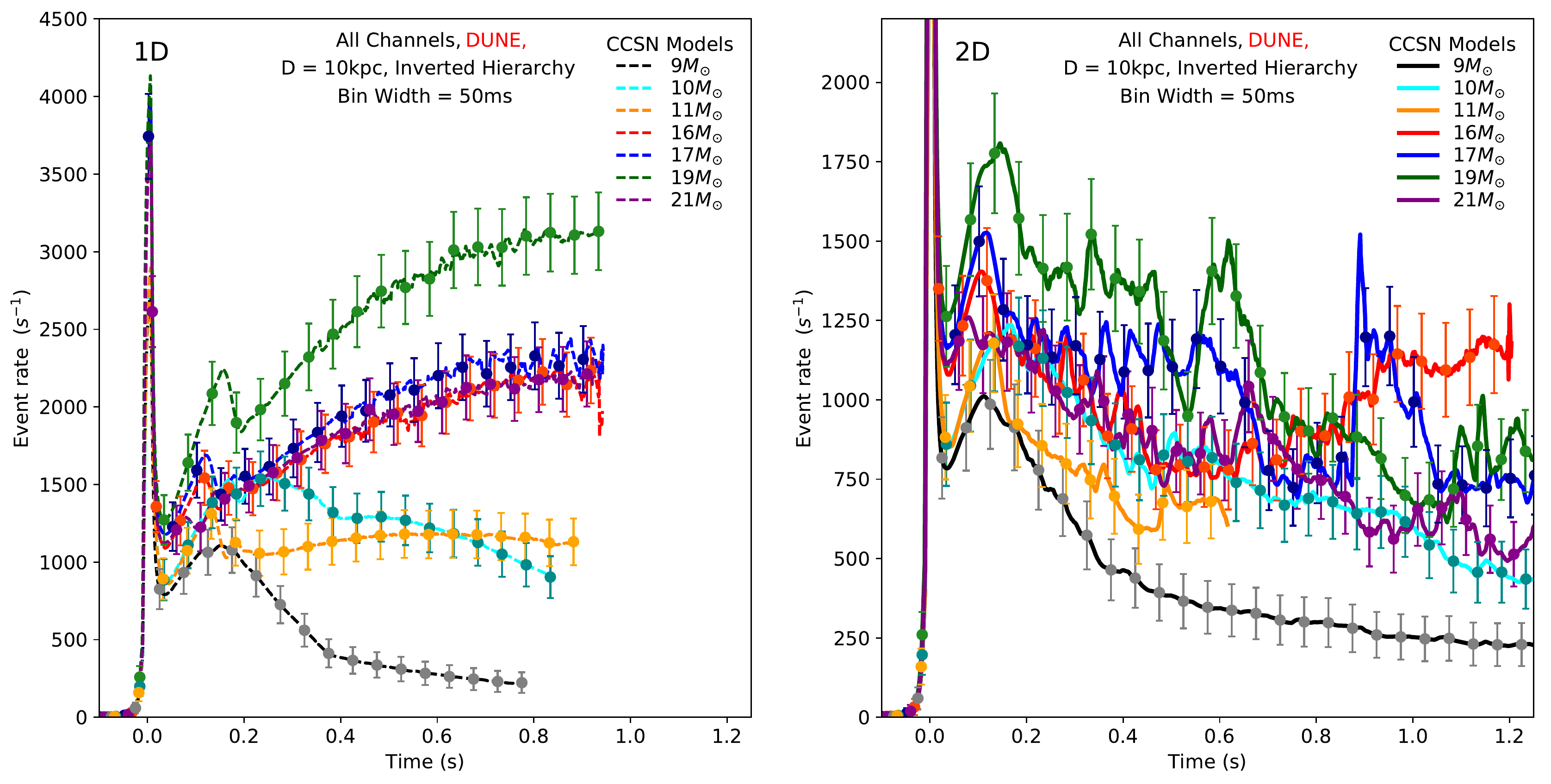}
\caption{Same as Fig. \ref{fig13}, but for DUNE.
\label{fig14}}
\end{figure}

Doing a full MCMC or Bayesian analysis to determine significance is beyond
the scope of this paper, which has, as discussed, much more limited ambitions.
However, we provide in Figures \ref{fig13} and \ref{fig14} neutrino light curves that include Poissonian
error bars at 10 kiloparsecs and for time bins of 50 milliseconds for 
Super-K and DUNE for the suite of mass models from 9 to 21 solar masses 
and in the NH and IH contexts.  This enables one to gauge the ability 
of these detectors to separate out the various masses.  As can be seen from 
Figures \ref{fig13} and \ref{fig14}, the model discrimination capability of these detectors
interior to $\sim$10 kiloparsecs is rather good.  Time binning is at the discretion
of those performing the data analysis, so had we used in Figures \ref{fig13} and \ref{fig14} 
even coarser binning the capacity to discriminate models within our galaxy 
would have been demonstrated to be even better.

In Table \ref{tab:PPer2}, we provide the ranges for Super-K, JUNO, and DUNE at which
the signal-to-noise ratio is five for a measurement of the flux at the first 
post-bounce peak, centered in a 10-ms bin around that time. We do this for 
all models and for both the Normal and Inverted hierarchies. This procedure can be
applied to any light-curve feature, and the results depicted in 
Table \ref{tab:PPer2} are merely representative. Nevertheless, Table \ref{tab:PPer2}
gives the reader a quantitative estimate of the supernova ranges out to which an important
light-curve feature could be discerned by an important subset of current and 
near-term underground neutrino observatories.

\subsection{Temporal Feature Registration}
\label{accuracy}

If a measurement is signal Poisson-limited (as it is for Super-K, DUNE and JUNO), the 
precision with which a temporal feature (a hump or fluctuation) can be registered in time goes
roughly as the feature width, $\Delta{t}$, divided by the square root of the event number, $N_i$, under the feature.
This is the ``centroiding" problem and implies that the distance out to which one can time-register 
a feature scales with the feature width\footnote{This is a very approximate statement of the more 
general Cramer-Rao theorem (Hogg \& Craig 1978).}. If the feature has a width of 10 milliseconds, 
it will require $N_i \sim 100$ to register the feature to within $\sim$1 millisecond. As Wallace et al. (2016)
have shown, to determine the peak time of the $\nu_e$ breakout burst to within $\sim$1 millisecond
in the inverted hierarchy requires a supernova at a distance of less than $\sim$4 kpc.  For the normal hierarchy, the 
breakout burst is much less easily captured (see \S\ref{osc}).  However, for a given underlying
model, the huge number of events expected during the subsequent dynamical phase would enable better
fractional registration of longer duration features.  For instance, using Super-K data, centroiding the accretion 
hump of $\sim$200 millisecond duration for the exploding 19-M$_{\odot}$ progenitor to within $\sim$10 milliseconds 
could in principle be done out to a range of $\sim$10 kpc.  For context, the mean time between single events 
during the SN 1987A campaign was $\sim$1 second.

We note that when a black hole forms, the neutrino signal should immediately shut off (Burrows 1984). 
This would be a precipitous drop, effectively an edge, and a detector's capacity to determine the precise time 
of black hole formation will be limited by the instantaneous event rate just before that time. 
Though for this paper we did not simulate the neutrino light curve for black hole formation,  the event
rates provided for the 1D models in the relevant figures are indicative of the approximate rates when 
such a secondary collapse to a black hole would occur.  Using these numbers suggests that the time of black
hole formation could be determined in Super-K, DUNE, or JUNO to one millisecond out to 10$-$20 kpc.

\section{Conclusions}
\label{conclusions}

Most of the neutrino burst signal diagnostic of supernova dynamics and the hydrodynamics 
of core-collapse occurs in this first $\sim$second after core bounce.  However, 
most of the signal occurs later during the subsequent tens-of-seconds of PNS cooling and 
deleptonization (Burrows \& Lattimer 1986).  Nevertheless, the neutrino light curve,
spectral evolution, and species mix during this first crucial phase uniquely 
bear the stamp of the internal dynamics of the supernova. In this paper, 
we have presented models of the all-channel neutrino light curve (event-rate/signal evolution) 
in four current and planned underground neutrino detectors for seven exploding and 
non-exploding progenitor models, with and without neutrino oscillations, in order to quantify 
the systematic variations of the neutrino event rate evolution with progenitor mass, 
oscillation model, and whether the model exploded. We identify features in the light 
curves diagnostic of each, as well as of the pre-collapse core mass density profile, the accretion of 
the silicon/oxygen interface, and temporal variations due to episodic accretion events. 
The network of large-volume neutrino detectors that is emerging around the world
will be hugely capable of constraining, perhaps in detail, the evolution in real time
of the supernova phenomenon for any galactic core-collapse supernova and providing
much-needed ground truth for the theory of supernovae upon which so many have labored
these last five decades\footnote{All of the models calculated in this paper are available upon request.}.

\section*{Acknowledgements}

The authors acknowledge support under U.S. NSF Grant AST-1714267 and  
the Max-Planck/Princeton Center (MPPC) for Plasma Physics (NSF PHY-1144374),  
and by the DOE SciDAC4 Grant DE-SC0018297 (subaward 00009650).
DR acknowledges support as a Frank and Peggy Taplin Fellow 
at the Institute for Advanced Study.
In addition, this material is based upon work supported by the 
U.S. Department of Energy, Office of Science, Office of Advanced 
Scientific Computing Research, and the Scientific Discovery through 
Advanced Computing (SciDAC) program under Award Number DE-SC0018297 (subaward 00009650).
The authors employed computational resources provided by the TIGRESS
high performance computer center at Princeton University, which is jointly supported by the Princeton
Institute for Computational Science and Engineering (PICSciE) and the Princeton University Office of
Information Technology, and by the National Energy Research Scientific Computing Center
(NERSC), which is supported by the Office of Science of the US Department of
Energy (DOE) under contract DE-AC03-76SF00098. The authors express their gratitude to
Ted Barnes of the DOE Office of Nuclear Physics for facilitating their use of NERSC.
%



\vfill\eject

\begin{landscape}
\begin{table}
\caption{Total number of events for a given progenitor for various detectors at 10 kiloparsecs during the
earlier ``supernova" dynamical phase. We provide the total number of accumulated events
within our calculational time near $\sim$1 second, starting at 50 ms before bounce
and ending near as long as $\sim$950 ms after bounce. These terminal times after
bounce are given in the table heading. NO stands for ``no oscillation", NH stands
for ``normal hierarchy", and IH stands for ``inverted hierarchy".} \centering
\begin{tabular}{c|c|c|c|c|c|c|c|c|c|c|c|c|c|c} \hline
\hline
 & \multicolumn{2}{c}{9 M$_\odot$} &\multicolumn{2}{c}{10 M$_\odot$} & \multicolumn{2}{c}{11 M$_\odot$} & \multicolumn{2}{c}{16 M$_\odot$} & \multicolumn{2}{c}{17 M$_\odot$} & \multicolumn{2}{c}{19 M$_\odot$} & \multicolumn{2}{c}{21 M$_\odot$} \\
 & 1D & 2D  & 1D & 2D  & 1D & 2D  & 1D & 2D  & 1D & 2D  & 1D & 2D  & 1D & 2D \\
 & (778 ms) & (950 ms)& (838 ms) & (950 ms)& (887 ms) & (616 ms) & (942 ms) & (950 ms)& (944 ms) & (950 ms)& (937 ms) &(950 ms) & (928 ms) & (950 ms)\\
 \hline

Super-K (NO) & 1057 & 1143 & 2114 & 1743 & 2100 & 1163 & 3449 & 2092 & 3683 & 2278 & 4652 & 2573 & 3402 & 1971  \\
Super-K (NH) & 880 & 1096 & 1600 & 1591 & 1401 & 1002 & 2184 & 1716 & 2321 & 1870 & 2837 & 2112 & 2147 & 1642   \\
Super-K (IH) &  815 & 1076 & 1452 & 1528 & 1223 & 947 & 1879 & 1585 & 1996 & 1724 & 2431 & 1944 & 1844 & 1526   \\
\hline 
DUNE (NO) & 471 & 476 & 1200 & 817 & 1244 & 534 & 2182 & 1011 & 2345 & 1132 & 3060 & 1275 & 2155 & 949  \\
DUNE (NH) & 485 & 620 & 897 & 929 & 711 & 552 & 1099 & 946 & 1168 & 1048 & 1413 & 1193 & 1077 & 910    \\
DUNE (IH) &  475 & 529 & 1083 & 859 & 1041 & 542 & 1762 & 987 & 1888 & 1102 & 2417 & 1247 & 1737 & 936   \\
\hline 
JUNO(NO) & 852 & 933 & 1652 & 1423 & 1643 & 948 & 2662 & 1701 & 2837 & 1852 & 3550 & 2093 & 2625 & 1611 \\
JUNO (NH)&   715 & 906 & 1249 & 1311 & 1096 & 827 & 1672 & 1409 & 1772 & 1534 & 2136 & 1732 & 1643 & 1357     \\
JUNO (IH) &  661 & 885 & 1128 & 1255 & 950 & 779 & 1425 & 1300 & 1508 & 1414 & 1804 & 1594 & 1398 & 1259   \\
\hline 
IceCube (NO) & 113854 & 123084 & 227649 & 187727 & 226195 & 125254 & 371439 & 225329 & 396647 & 245295 & 501013 & 277047 & 366332 & 212280    \\
IceCube (NH) &  94815 & 118007 & 172308 & 171384 & 150855 & 107922 & 235201 & 184820 & 249963 & 201397 & 305563 & 227457 & 231191 & 176868\\
IceCube (IH) &  87733 & 115928 & 156345 & 164540 & 131685 & 102004 & 202312 & 170721 & 215008 & 185686 & 261827 & 209335 & 198581 & 164353   \\

\hline 
\end{tabular}
\label{tab:PPer}
\end{table}
\end{landscape}

\begin{landscape}
\begin{table}
\caption{Ranges (in kpc) for the determination at 5-$\sigma$ of the neutrino event rate for a 10-ms bin centered
on the first peak following breakout for both the Normal and Inverted hierarchies. This is usually associated with the accretion-powered phase, but for DUNE, given that
the signal in it is dominated by the $\nu_e$ component at Earth, the peak times for a give model do not
match those in the corresponding neutrino light curves in Super-K, JUNO, and IceCube.  Times
at which the bins are centered for Super-K and JUNO are given in the heading and for DUNE in the body of the Table.} 
\centering
\begin{tabular}{c|c|c|c|c|c|c|c|c|c|c|c|c|c|c} \hline
\hline
 & \multicolumn{2}{c}{9 M$_\odot$} &\multicolumn{2}{c}{10 M$_\odot$} & \multicolumn{2}{c}{11 M$_\odot$} & \multicolumn{2}{c}{16 M$_\odot$} & \multicolumn{2}{c}{17 M$_\odot$} & \multicolumn{2}{c}{19 M$_\odot$} & \multicolumn{2}{c}{21 M$_\odot$} \\
 & 1D & 2D  & 1D & 2D  & 1D & 2D  & 1D & 2D  & 1D & 2D  & 1D & 2D  & 1D & 2D \\
 & (109 ms) & (101 ms)& (164 ms) & (124 ms)& (131 ms) & (113 ms) & (112 ms) & (96 ms)& (118 ms) & (103 ms)& (153 ms) &(105 ms) & (76 ms) & (76 ms)\\
 \hline

Super-K (NH) & 9.60 & 9.35 & 10.61 & 10.01 & 10.27 & 9.90 & 11.25 & 10.88 & 11.62 & 11.20 & 12.51 & 11.81 & 10.91 & 10.73  \\
Super-K (IH) &  9.01 & 8.83 & 9.93 & 9.47 & 9.58 & 9.30 & 10.52 & 10.26 & 10.86 & 10.55 & 11.59 & 11.10 & 10.30 & 10.18  \\
\hline 
JUNO (NH) & 8.70 & 8.48 & 9.59 & 9.09 & 9.30 & 8.98 & 10.19 & 9.88 & 10.53 & 10.16 & 11.26 & 10.71 & 9.89 & 9.74    \\
JUNO (IH) &  8.16 & 8.00 & 8.98 & 8.59 & 8.68 & 8.43 & 9.53 & 9.30 & 9.83 & 9.57 & 10.45 & 10.06 & 9.33 & 9.22   \\
\hline 
 & (113 ms) & (112 ms)& (170 ms) & (116ms)& (134 ms) & (114 ms) & (115 ms) & (106 ms)& (120 ms) & (110 ms)& (157 ms) &(112 ms) & (79 ms) & (81 ms)\\
DUNE (NH) & 6.89 & 6.69 & 7.85 & 7.23& 7.43 & 7.11 & 8.20 & 7.90 & 8.51 & 8.15 & 9.28 & 8.628 & 7.88 & 7.75   \\
& (153 ms) & (112 ms)& (170 ms) & (168 ms)& (134 ms) & (127 ms) & (115 ms) & (106 ms)& (120 ms) & (118 ms)& (157 ms) &(145 ms) & (79 ms) & (81 ms)\\
DUNE (IH) &  6.65 & 6.36 & 7.83 & 7.03 & 7.25 & 6.88 & 7.86 & 7.50 & 8.22 & 7.81 & 9.46 & 8.50 & 7.18 & 7.10   \\

\hline 
\end{tabular}
\label{tab:PPer2}
\end{table}
\end{landscape}

\onecolumn

\onecolumn


\bsp	
\label{lastpage}
\end{document}